\documentclass[twoside,twocolumn]{article}

\newlength{\irrl}
\newlength{\irrw}

\newcommand{\n}{\bm{n}}
\newcommand{\normal}{\bm{\nu}}
\newcommand{\Id}{\mathbf{I}}
\newcommand{\Ws}{W_\mathrm{s}}
\newcommand{\Wa}{W_\mathrm{a}}
\newcommand{\e}{\bm{e}}
\newcommand{\ex}{\e_x}
\newcommand{\ey}{\e_y}
\newcommand{\ez}{\e_z}
\newcommand{\mol}{\bm{m}}
\newcommand{\twist}{\bm{t}}
\renewcommand{\frame}{(\e_1,\e_2)}
\newcommand{\surface}{\mathscr{S}}
\newcommand{\curve}{\mathscr{C}}
\newcommand{\ca}{\vartheta}

\newcommand{\TBN}{$\mathrm{N}_\mathrm{tb}$\ }

\newcommand{\az}{\varphi}
\newcommand{\xii}{\xi_\infty}
\newcommand{\ens}{\mathscr{F}_\mathrm{s}}
\newcommand{\Pt}{\Id-\twist\otimes\twist}
\newcommand{\ave}[1]{\left\langle{#1}\right\rangle}
\newcommand{\jump}[1]{\mbox{$\left\llbracket #1\right\rrbracket$}}
\newcommand{\oone}{\omega_\mathrm{c}^{(1)}}
\newcommand{\otwo}{\omega_\mathrm{c}^{(2)}}
\newcommand{\uv}{\bm{u}}
\newcommand{\rv}{\bm{r}}
\newcommand{\vae}{\varepsilon}
\newcommand{\tp}{\bm{t}^+}
\newcommand{\tm}{\bm{t}^-}
\newcommand{\sqc}{\sqrt{1-c^2}}
\newcommand{\f}{\bm{f}}
\newcommand{\sepa}{\frac{1}{\sqrt{2}}}
\newcommand{\Tm}{$T_\mathrm{m}$}
\newcommand{\Ts}{$T_\mathrm{s}$}
\newcommand{\degree}{^\circ}

\usepackage{amsmath}
\usepackage{amsfonts}
\usepackage{bm}
\usepackage{bbm}
\usepackage{amssymb}
\usepackage{mathrsfs}
\usepackage{dsfont}
\usepackage{subfigure}
\usepackage{paralist}
\usepackage{wasysym}
\usepackage{stmaryrd}
\usepackage{verbatim}
\usepackage[super,sort&compress,comma]{natbib}
\usepackage{mhchem}
\usepackage{times}
\usepackage{sectsty}
\usepackage{balance}

\usepackage{graphicx} 
\usepackage{lastpage}
\usepackage{fancyhdr}
\usepackage{lastpage}
\usepackage{float}
\usepackage{fancyhdr}
\usepackage{fnpos}
\usepackage{array}
\usepackage{charter}
\usepackage[T1]{fontenc}
\usepackage[usenames,dvipsnames]{xcolor}
\usepackage{setspace}
\usepackage[compact]{titlesec}

\usepackage{epstopdf}

\definecolor{cream}{RGB}{222,217,201}

\begin{document}

\pagestyle{fancy}
\thispagestyle{plain}
\fancypagestyle{plain}{

\fancyhead[C]{\includegraphics[width=18.5cm]{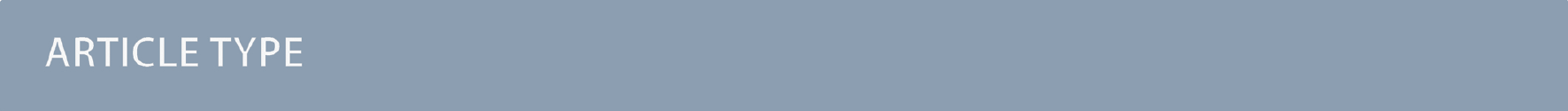}}
\fancyhead[L]{\hspace{0cm}\vspace{1.5cm}\includegraphics[height=30pt]{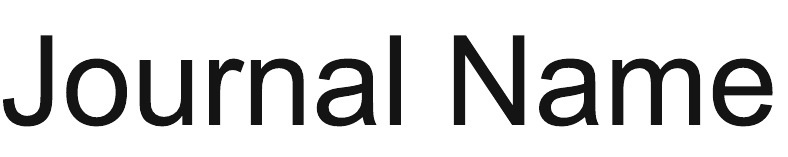}}
\fancyhead[R]{\hspace{0cm}\vspace{1.7cm}\includegraphics[height=55pt]{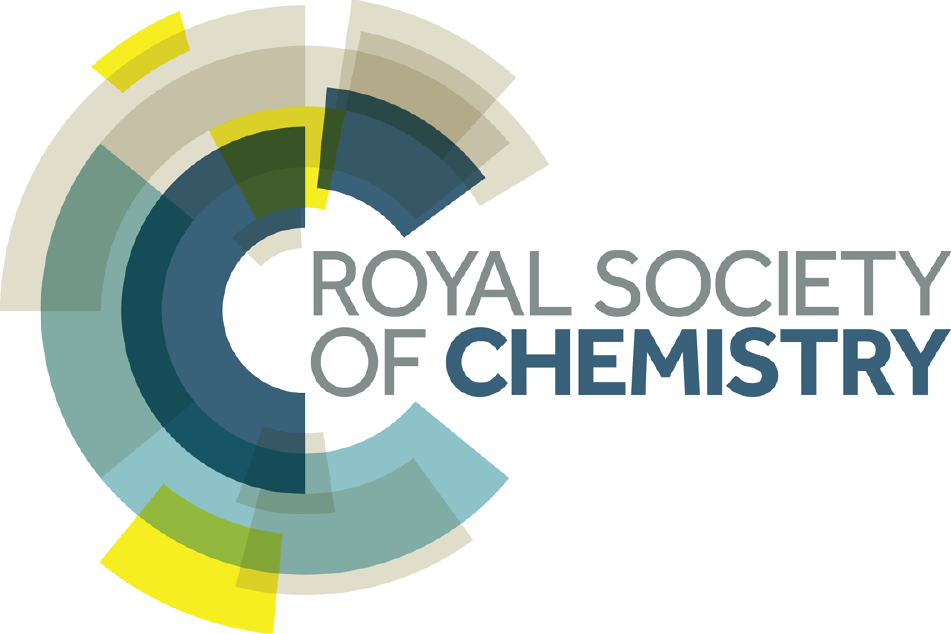}}
\renewcommand{\headrulewidth}{0pt}
}

\makeFNbottom
\makeatletter
\renewcommand\LARGE{\@setfontsize\LARGE{15pt}{17}}
\renewcommand\Large{\@setfontsize\Large{12pt}{14}}
\renewcommand\large{\@setfontsize\large{10pt}{12}}
\renewcommand\footnotesize{\@setfontsize\footnotesize{7pt}{10}}
\makeatother

\renewcommand{\thefootnote}{\fnsymbol{footnote}}
\renewcommand\footnoterule{\vspace*{1pt}%
\color{cream}\hrule width 3.5in height 0.4pt \color{black}\vspace*{5pt}}
\setcounter{secnumdepth}{5}

\makeatletter
\renewcommand\@biblabel[1]{#1}
\renewcommand\@makefntext[1]%
{\noindent\makebox[0pt][r]{\@thefnmark\,}#1}
\makeatother
\renewcommand{\figurename}{\small{Fig.}~}
\sectionfont{\sffamily\Large}
\subsectionfont{\normalsize}
\subsubsectionfont{\bf}
\setstretch{1.125} 
\setlength{\skip\footins}{0.8cm}
\setlength{\footnotesep}{0.25cm}
\setlength{\jot}{10pt}
\titlespacing*{\section}{0pt}{4pt}{4pt}
\titlespacing*{\subsection}{0pt}{15pt}{1pt}

\fancyfoot{}
\fancyfoot[LO,RE]{\vspace{-7.1pt}\includegraphics[height=9pt]{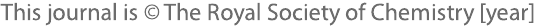}}
\fancyfoot[CO]{\vspace{-7.1pt}\hspace{13.2cm}\includegraphics{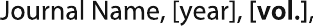}}
\fancyfoot[CE]{\vspace{-7.2pt}\hspace{-14.2cm}\includegraphics{RF}}
\fancyfoot[RO]{\footnotesize{\sffamily{1--\pageref{LastPage} ~\textbar  \hspace{2pt}\thepage}}}
\fancyfoot[LE]{\footnotesize{\sffamily{\thepage~\textbar\hspace{3.45cm} 1--\pageref{LastPage}}}}
\fancyhead{}
\renewcommand{\headrulewidth}{0pt}
\renewcommand{\footrulewidth}{0pt}
\setlength{\arrayrulewidth}{1pt}
\setlength{\columnsep}{6.5mm}
\setlength\bibsep{1pt}

\makeatletter
\newlength{\figrulesep}
\setlength{\figrulesep}{0.5\textfloatsep}

\newcommand{\topfigrule}{\vspace*{-1pt}%
\noindent{\color{cream}\rule[-\figrulesep]{\columnwidth}{1.5pt}} }

\newcommand{\botfigrule}{\vspace*{-2pt}%
\noindent{\color{cream}\rule[\figrulesep]{\columnwidth}{1.5pt}} }

\newcommand{\dblfigrule}{\vspace*{-1pt}%
\noindent{\color{cream}\rule[-\figrulesep]{\textwidth}{1.5pt}} }

\makeatother

\twocolumn[
  \begin{@twocolumnfalse}
\vspace{3cm}
\sffamily
\begin{tabular}{m{4.5cm} p{13.5cm} }

\includegraphics{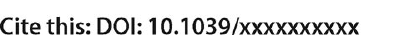}
& \noindent\LARGE{\textbf{Interfacial and morphological features of a twist-bend nematic drop}} \\
\vspace{0.3cm} & \vspace{0.3cm} \\

& \noindent\large{Kanakapura S. Krishnamurthy,\textit{$^a$} Pramoda Kumar,\textit{$^b$} Nani B. Palakurthy,\textit{$^a$} Channabasaveshwar V. Yelamaggad,\textit{$^a$} and Epifanio G. Virga$^\ast$\textit{$^c$}} \\

\includegraphics{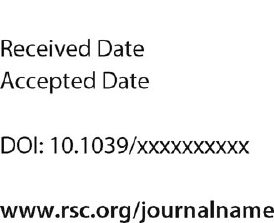}
& \noindent\normalsize{In this experimental and theoretical study, we examine the equilibrium shapes of quasi-two-dimensional twist-bend nematic ($\mathrm{N_{tb}}$) drops formed within a planarly aligned nematic layer of the liquid crystal CB7CB. Initially, at the setting point of the \TBN phase, the drops assume a nonequilibrium cusped elliptical geometry with the major axis orthogonal to the director of the surrounding nematic fluid; this growth is governed principally by anisotropic heat diffusion. The drops attain equilibrium through thermally driven dynamical evolutions close to their melting temperature. They are associated with a characteristic twin-striped morphology that transforms into the familiar focal conic texture as the temperature is lowered. At equilibrium, large millimetric drops are tactoidlike, elongated along the director of the surrounding nematic fluid. This geometry is explained by a mathematical model that features two dimensionless parameters, of which one is the structural cone angle of the \TBN phase and the other, the relative strength of mismatch elastic energy at the drop's interface. Both parameters are extracted from the observations by measuring the aspect ratio of the equilibrium shapes and the inner corner angle of the cusps.} \\

\end{tabular}

 \end{@twocolumnfalse} \vspace{0.6cm}

 ]


\renewcommand*\rmdefault{bch}\normalfont\upshape
\rmfamily
\section*{}
\vspace{-1cm}


\footnotetext{\textit{$^{a}$~Centre for Nano and Soft Matter Sciences, P. O. Box 1329, Jalahalli, Bangalore 560013, India. E-mail: murthyksk@gmail.com}}
\footnotetext{\textit{$^{b}$~The Jacob Blaustein Institutes for Desert Research, Ben-Gurion University of the Negev, Sede Boqer Campus, Israel 8499000.}}
\footnotetext{\textit{$^{c}$~Dipartimento di Matematica, Universit\`a di Pavia, Via Ferrata 5, I-27100 Pavia, Italy. E-mail: eg.virga@unipv.it}}

\section{Introduction}\label{sec:intro}
The Wulff construction that relates the equilibrium shape of a crystalline particle to the specific surface free energy of its polyhedral faces\cite{wulff:frage,herring:some} has long been known to be extendable to liquid crystal drops, whether they are faceted or not.\cite{herring:use,chandrasekhar:surface,virga:variational}  For instance, the geometry of minimal surface of a small nematic (N) drop formed in its own isotropic (I) melt is obtained simply from a polar plot of the interfacial energy density.\cite{virga:drops} As a macroscopic expression of the orientational order in the drop, this shape turns out to be akin to an ellipsoid in general; when the molecules are rodlike (calamitic), the major axis of the shape lies along their preferred direction of orientation $\n$ (director).

For the particular case when the anisotropy of interfacial energy is large, the figure is modified into a \emph{tactoid} with singularities in the surface normal located at the extremities of the major axis. Structurally, the calamitic director $\n$ is usually tangential to the N-I interface, with the distortion being splay-rich near the vertices where point singularities (boojums) are formed, and bend-rich near the equator.  Spindle-shaped formations with nematic order are obtained typically in lyotropic systems. They were observed as early as 1925 in aqueous colloidal suspensions of vanadium pentoxide.\cite{zocher:freiwillige,zocher:taktosole}  A recent model of prolate tactoids in \ce{V2O5}-water systems takes into account the elastic and surface energies, together with the interaction energy between the director field and the boundary of the tactoid; it relates the tactoid-shape to the ratio of bend elastic constant $K_3$ to splay elastic constant $K_1$.\cite{kaznacheev:nature,kaznacheev:influence}

Tactoids have also been found in many macromolecular systems. Bernal and Frankuchen\cite{bernal:x-ray} found nematic tactoids in isotropic background as well as tactoidal isotropic inclusions in preparations of tobacco mosaic virus dispersed in water. More recently, tactoids in F-actin solutions were observed to form either through nucleation and growth or via spinodal decomposition.\cite{oakes:growth} Lately, extensive studies have been conducted on nematic tactoids in lyotropic chromonic liquid crystals.\cite{tortora:self-assembly,tortora:chiral,kim:morphogenesis,peng:chirality,jeonga:chiral,mushenheim:using}
There have also been several theoretical results concerning the structure and stability of lyotropic nematic drops.\cite{prinsen:shape,prinsen:continuous,prinsen:parity} For instance, it is predicted that virtual point defects located outside the poles of a tactoid approach each other with increasing droplet size to become true boojums only in the limit of infinite volume.\cite{prinsen:continuous}  Similarly, it is shown that, under certain elastic anisotropy criteria, the director field in a N-tactoid acquires a twisted, parity-broken structure.\cite{prinsen:parity}

Morphology of N-drops has also been investigated in a variety of thermotropic systems. For example, N-tactoids have been found to form within smectic membranes of a thermotropic liquid crystal.\cite{dolganov:shape}  N-droplets that appear in the course of N-N phase separation in mixtures of calamitic and polymeric mesogens are found to be spindle-shaped and, interestingly, this geometry is here attributed to a very low interfacial tension.\cite{casagrande:observation} In a study of the N-SmB (smectic B) interface in a quasi-two-dimensional sample, N germs forming in the background of the SmB phase have been found to be non-faceted and oval shaped; on the other hand, planar SmB germs in either planar or homeotropic N environment are of rounded rectangular geometry.\cite{buka:equilibrium} For two-dimensional (2D) materials with only orientational order and no positional order, cusped shapes are, in fact, considered to be a generic feature\cite{rudnick:shape,rudnick:theory} along with their regularized counterparts (also called \emph{dips} and \emph{quasicusps}).\cite{galatola:new}   As discussed later, it is of some relevance to the present study that the phase diagram depicting typical 2D nematic domains in the anchoring strength--surface tension space includes a regime of near circular domains with bipolar field and rounded tips.\cite{bijnen:texture}

In the background of foregoing results on the morphology of orientationally ordered mesomorphic domains, we may now consider the title subject.  The twist-bend nematic ($\mathrm{N_{tb}}$) phase, also described as \emph{heliconical}, has its ground state with achiral molecules disposed in a helical manner around an axis referred to as the twist director $\twist$, with a pitch of the order of $10\,\mathrm{nm}$.\cite{dozov:spontaneous}  The local preferred orientation of molecules $\mol$, unlike in a cholesteric where it is orthogonal to $\twist$, is oblique relative to $\twist$ (by around $20$-$30$ deg.). Despite its resemblance to the chiral smectic phase (SmC*), the mass density in the \TBN is uniformly distributed in space and the planes with uniform molecular alignment do not define material layers; to mark the distinction, it has been proposed to call the latter \emph{pseudo-layers}.\cite{challa:twist-bend} Molecules move freely from one pseudo-layer to the next, constituting a truly three-dimensional, anisotropic liquid. The \TBN phase, which is perhaps the most intriguing of lately discovered mesophases, is currently of immense research interest. \cite{panov:spontaneous,meyer:flexoelectrically,borshch:nematic,xiang:electrically,virga:double-well,hernandez:twist} It is the purpose here to dwell on our first experimental and theoretical exploration of the morphological features of \TBN drops forming under cooling in the nematic background. We have observed tactoid-like drops both on micrometer and millimeter scales, under quasi-equilibrium conditions.  Non-equilibrium geometry of the drops is nearly elliptical involving vertices at the extremities of the principal axes; it is realized under relatively large undercooling and is dominated by kinetic factors.

We propose to present and discuss these results in two parts, the first dealing with experimental observations and the second, with the theoretical description based on a coarse-graining model. More specifically, in Sec.~\ref{sec:experiments}, we describe the experimental set up and we present the experimental findings. Section \ref{sec:theory} is devoted to a mathematical model for a 2D \TBN drop surrounded by its nematic phase, which builds on a coarse-grained surface energy for the interface between the two differently ordered phases. Finally, in Sec.~\ref{sec:conclusion} we summarize the main results of this paper and comment on their possible significance in the growing body of studies on \TBN phases. The paper is supplemented by an Appendix containing the analytical details of the mathematical model employed in Sec.~\ref{sec:two_dimensional_drops}.

\section{Experiments}\label{sec:experiments}
Morphological and interfacial features of the \TBN phase were studied in cyanobiphenyl-\ce{(CH2)7}-cyanobiphenyl (CB7CB) exhibiting the phase sequence: $\mathrm{N_{tb}}\text{---}103.3\,\degree\mathrm{C}$(\Tm)$\rightarrow\text{N}\text{---}116.5\,\degree\mathrm{C}\rightarrow\text{Isotropic}$ (heating); $\text{Isotropic---}116\,\degree\mathrm{C}\rightarrow\text{N}\text{---}103\,\degree\mathrm{C}$(\Ts)$\rightarrow\mathrm{N_{tb}}$ (cooling); here \Tm\ and \Ts\ denote the \emph{melting} and \emph{setting} temperatures of the \TBN phase, respectively.  The sample cells supplied by M/s AWAT, Poland were sandwich type, constructed of indium tin oxide coated glass plates. The electrodes were overlaid with polyimide and buffed unidirectionally to secure planar anchoring. The cell gap $d$ was either $9\,\mu\mathrm{m}$ or $20\,\mu\mathrm{m}$. Optical textures were examined using a Carl-Zeiss Axio Imager.M1m polarizing microscope equipped with an AxioCam MRc5 digital camera. The sample temperature $T$ was maintained to an accuracy of $\pm0.1\,\degree\mathrm{C}$ by an Instec HCS402 hot-stage connected to a STC200 temperature controller. The viewing aperture of the stage was $5\,\mathrm{mm}$ in diameter and the temperature at its centre was slightly lower (by $\sim0.2\,\degree\mathrm{C}$) compared to the temperature in the outer isothermal region.  This ensured nucleation of the N phase within the isotropic liquid, and of the \TBN phase within the nematic fluid in the central region of the aperture. Our main findings relating to \TBN drops are in the temperature range of $0.3\,\degree\mathrm{C}$ between \Tm\ and \Ts. In this narrow range, considering that even the intensity of illumination of the sample could affect the actual $T$ in the visual field, the cited temperatures are to be treated as indicative of trends in morphological changes.

\subsection{Morphological features of N and \TBN drops}\label{sec:morphological}
Before taking up the \TBN phase separation, for later comparison, we may first describe the geometry of the nematic germ.  Upon a slow cooling of the isotropic liquid, the N phase nucleates at $T=116\,\degree\mathrm{C}$ as a defect-free monodomain with the molecular alignment along the rubbing
direction $x$. The domain, which is essentially two-dimensional (2D), possesses a nearly elliptical geometry in the sample plane (normal to the viewing direction $z$). This is the cusp-free Wulff shape predicted previously for the nematic drop under the assumption of a moderate anisotropy of interfacial energy.\cite{chandrasekhar:surface} Significantly, regardless of the area of the sample under view, the nucleation occurs in the central region of the hot-stage aperture. In other words, we have here a case of homogeneous nucleation taking place in a marginally undercooled nematic fluid, and this is indicative of a very low free energy barrier.
\begin{figure}[ht]
  \centering
  \includegraphics[width=.8\linewidth]{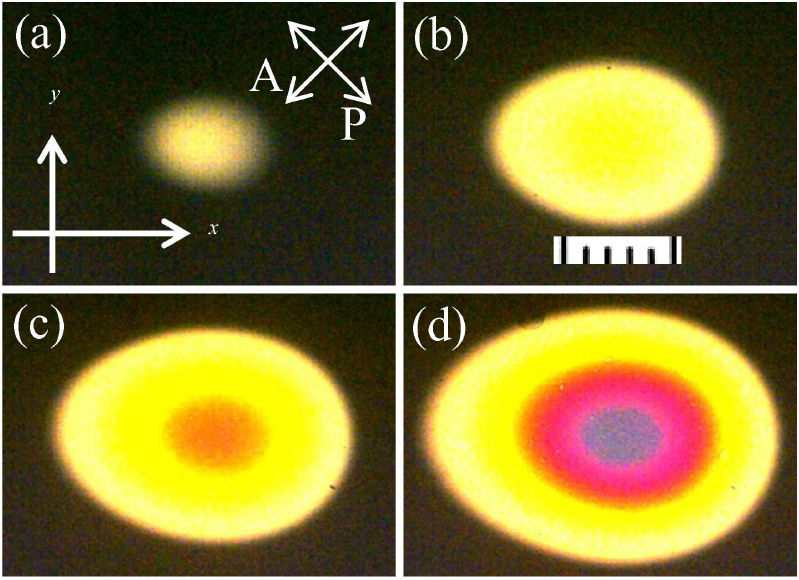}
  \caption{
  Homogeneous nucleation and growth of the N-phase in the form of an oval shaped droplet surrounded by the isotropic phase at $T\approx116\,\degree\mathrm{C}$. The director $\n$ in the drop is along the rubbing direction $x$.  Separation between successive panels is $\sim1.53\,\mathrm{s}$. Polarizer P and analyzer A are diagonally crossed.  $d=9\,\mu\mathrm{m}$; $0.1\,\mathrm{mm}$ each scale div.
  }
\label{fig:Fig.1}
\end{figure}
Figure~\ref{fig:Fig.1} shows some selected frames of a time series recorded during the growth of a nematic drop. The increase in interference colour at the centre, between Figs.~\ref{fig:Fig.1}a and \ref{fig:Fig.1}b, corresponds to a change in optical path by about $350\,\mathrm{nm}$ or an increase in birefringence by $0.039$.
\begin{figure}[ht]
  \centering
  \includegraphics[width=.8\linewidth]{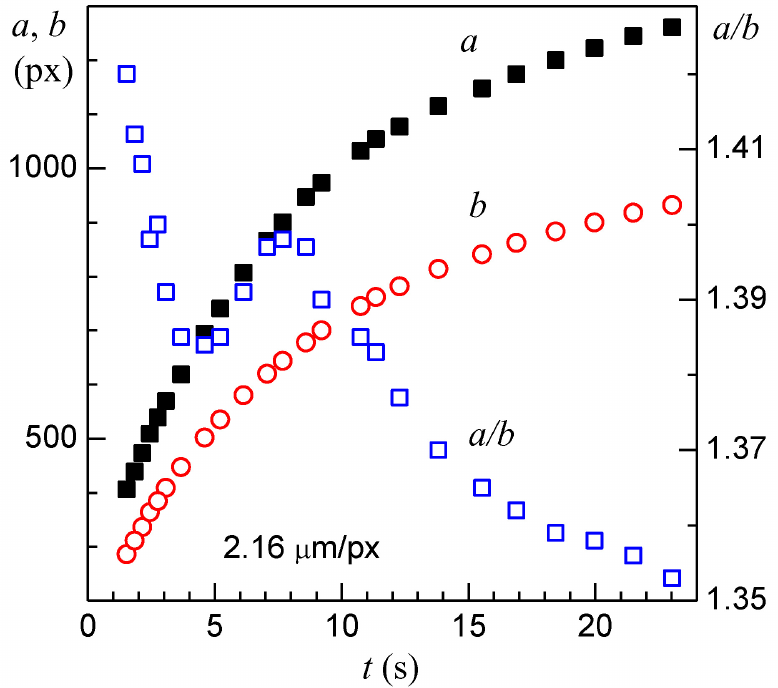}
  \caption{
  Growth of a nearly elliptical monodomain of the nematic liquid crystal within the isotropic phase of CB7CB held in a sandwich cell with $9\,\mu\mathrm{m}$ gap; $a$ and $b$ denote the major and minor axes of the domain.
  }
\label{fig:Fig.2}
\end{figure}
This is attributable to a very slight lowering of overall temperature during the growth. Figure~\ref{fig:Fig.2} depicts the time variation of the principal dimensions of the drop. While the growth rates of major and minor axes, $a$ and $b$, decrease with increasing time, the aspect ratio $a/b$ is nonmonotonous, showing an overall drop from $1.43$ to $1.35$ during the growth progresses.

\begin{figure}[ht]
  \centering
  \includegraphics[width=.8\linewidth]{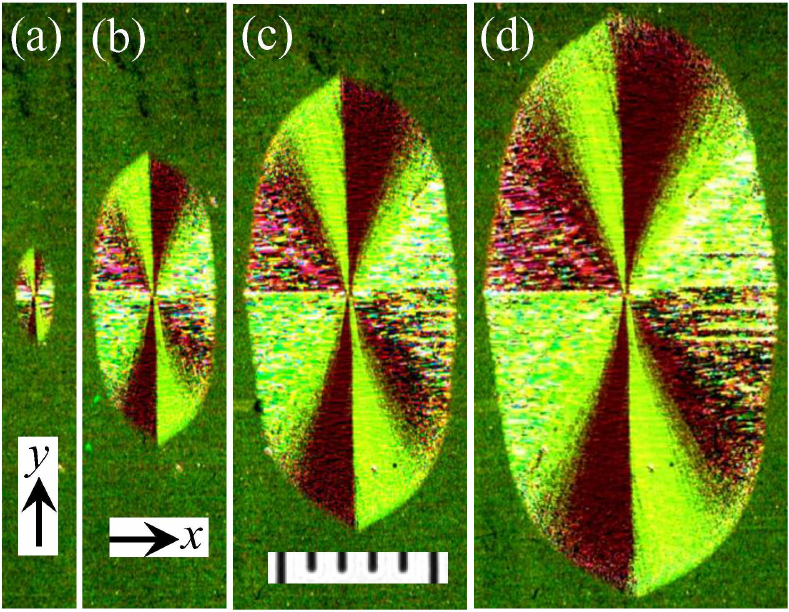}
  \caption{
  Twist-bend nematic LC drop growing in the nematic phase of CB7CB observed between polarizer at $45\degree$ to $x$ and analyzer along $y$. Nucleation occurred at the centre of the hot-stage circular window where the temperature was marginally lower than outside the window. Sample is well aligned in the nematic along the rubbing direction $x$. The \TBN has the helical axis largely along the rubbing direction, or along the short axis of the near-elliptic drop. The near elliptic drop has its principal axes terminating in vertices. Interval between successive frames is $0.302\,\mathrm{s}$. $d=9\,\mu\mathrm{m}$.   $0.1\,\mathrm{mm}$ each scale div.
  }
\label{fig:Fig.3}
\end{figure}

\begin{figure}[ht]
  \centering
  \includegraphics[width=.8\linewidth]{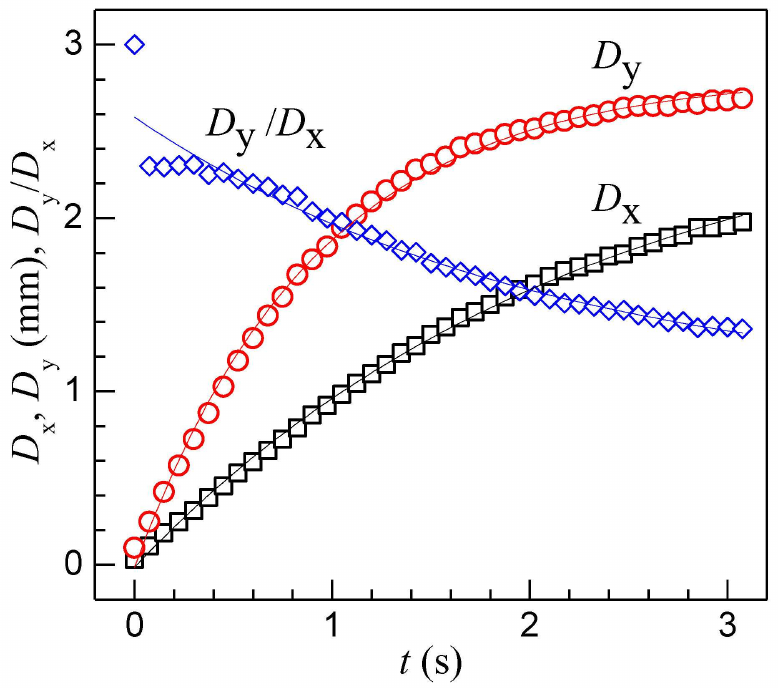}
  \caption{
  Growth of a nearly elliptical domain of the \TBN phase within the nematic phase of CB7CB held in a sandwich cell with $9\,\mu\mathrm{m}$ gap; $D_x$ and $D_y$ denote the dimensions along $x$ and $y$. Continuous curves are exponential fits. $T=103\,\degree\mathrm{C}$.
  }
\label{fig:Fig.4}
\end{figure}

The nematic phase of CB7CB supercools slightly, by $\sim0.3\,\degree\mathrm{C}$, before undergoing transition to the \TBN phase. The \TBN domain, which nucleates at $T_\mathrm{s}=103\,\degree\mathrm{C}$, displays a characteristic geometry
illustrated in Fig.~\ref{fig:Fig.3}.
It may be described as a cylindrulite, approximately elliptic in planform and extending predominantly along the normal to the rubbing direction $x$. In addition, cusp-like  formations are found at the opposite ends of the principal axes. The time dependence of principal dimensions $D_x$ and $D_y$, and their ratio $D_y/D_x$ is shown in Fig.~\ref{fig:Fig.4}.

\begin{figure}[ht]
  \centering
  \includegraphics[width=.8\linewidth]{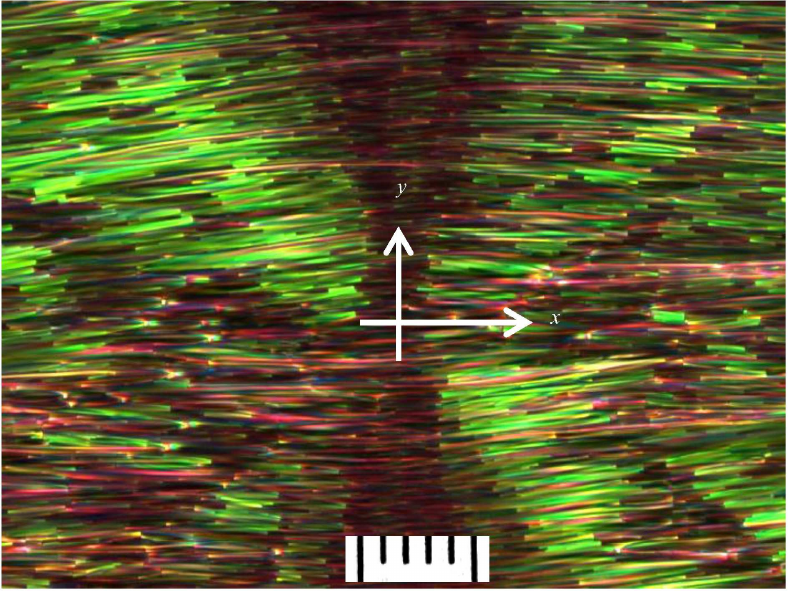}
  \caption{
  Central region of an \TBN drop showing the biconvex texture formed by the stripes. $20\,\mu\mathrm{m}$ each scale div. Crossed polarizers set along $(x, y)$. $T=102.9\,\degree\mathrm{C}$.
  }
\label{fig:Fig.5}
\end{figure}

\begin{figure}[ht]
  \centering
  \includegraphics[width=.8\linewidth]{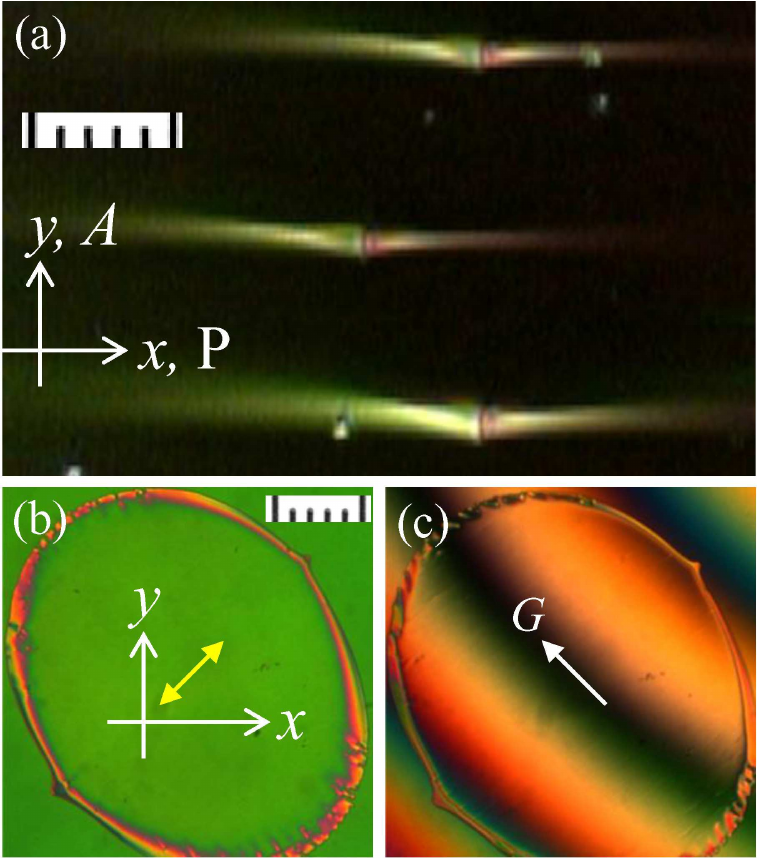}
  \caption{
  (a) Defect structures distinctive of the \TBN phase very close to the $\mathrm{N_{tb}}$-N transition temperature.  (b) A large, near-elliptical \TBN domain with its major axis oriented diagonally relative to $(x, y)$, as viewed between crossed polarizers, P (along $x$) and A (along $y$). The domain in the nonequilibrium state is slowly collapsing along its major axis relative to the minor axis joining the cusps. Within the domain, the interference colour is uniform with the structural defects causing no perceptible change in birefringence. (c) Tilt compensator with its slow axis along $\mathrm{G}$ is set for compensation in the mid region. Each scale div. is $4\,\mu\mathrm{m}$ in (a) and $100\,\mu\mathrm{m}$ in (b) and (c).}
\label{fig:Fig.6}
\end{figure}

Comparing the N and \TBN domains in Figs.~\ref{fig:Fig.1} and \ref{fig:Fig.3}, we notice that, apart from their extension along orthogonal directions, they differ fundamentally in terms of structural homogeneity. Unlike the former, which is defect-free, the latter is replete with defects that appear as stripes. A careful examination of the drop shows that the stripes tend to occur along curved lines on either side of the minor axis, leading to a tactoid-like formation with the vertices along $x$ (Fig.~\ref{fig:Fig.5}).
The number of stripe defects decreases rapidly as the sample temperature is increased from  \Ts\ to \Tm. An enlarged view of stripe defects existing within a nearly elliptical \TBN domain at $T=103.2\,\degree\mathrm{C}$ is presented in Fig.~\ref{fig:Fig.6}a. Each of the defects is composed of two bright stripes seen against a dark background, with an elliptical-looking disclination loop formed at the junction; the stripes differ in their birefringence colour, being inclined oppositely by a small angle relative to the rubbing direction.  The optical path corresponding to the stripes is about the same as that of the surrounding region. This is evident from Fig.~\ref{fig:Fig.6}b wherein the defects are invisible in the \TBN domain diagonally disposed relative to $(x, y)$.
Consequently, as evident in Fig.~\ref{fig:Fig.6}c, extinction of light occurs for the same position of the tilt compensator all along the diagonal band containing several stripe defects (visible on rotation of the stage by $45\degree$). It is not clear if these defects (Fig.~\ref{fig:Fig.6}a) are focal conic structures with the stripe direction coinciding with the twist axis in the far field. For the present, we may refer to them as \emph{twin-stripes}. Right at \Tm\ at which the \TBN phase coexists with the N phase, the stripe defects are not observed (see later).  Therefore, it is possible that the twist axis in the \TBN drop at \Tm\ is along the easy axis of the surrounding nematic; and with lowering $T$, stripe regions with inclined twist direction develop in increasing number. Alternatively, close to \Tm, twin-stripes, which are very narrow and optically unresolved, overlap so as to result in the effective optical axis along $x$; at lower temperatures they reorganize into large-size bands inclined to $x$.

\begin{figure*}[ht]
  \centering
  \includegraphics[width=.75\linewidth]{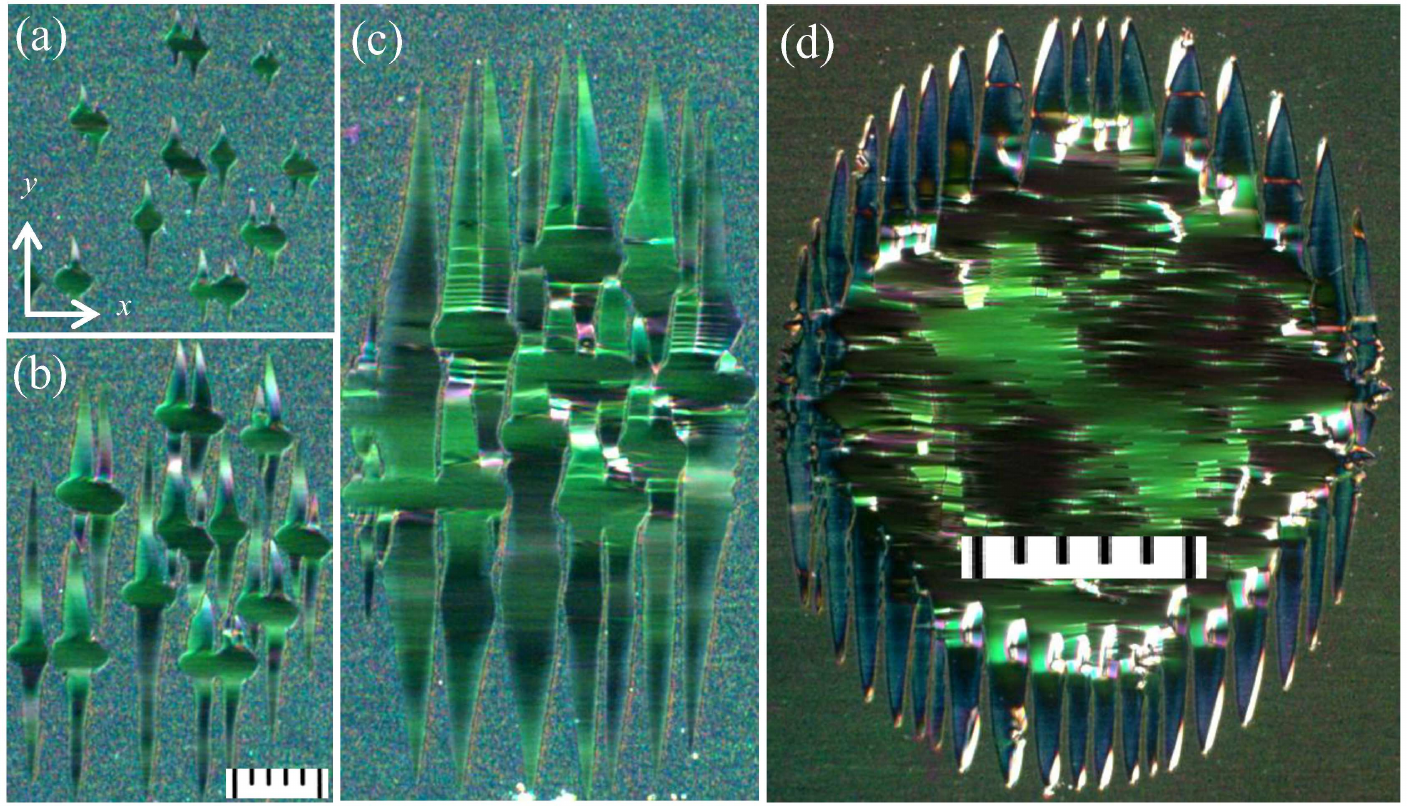}
  \caption{
  Directed growth of \TBN fingers along $\pm y$ direction. (a)-(c) Frames from a time series with (b) and (c) captured, relative to (a), after $2.7\,\mathrm{s}$ and $8.3\,\mathrm{s}$, respectively; $20\,\mu\mathrm{m}$ each scale div. (d) Image from a different series showing a large \TBN domain within which are formed numerous stripe defects, all oriented approximately along $x$; $0.1\,\mathrm{mm}$ each scale div. Polarizer along $x$ and analyzer nearly crossed.}
\label{fig:Fig.7}
\end{figure*}

The reason for the shape of the \TBN domain such as seen in Fig.~\ref{fig:Fig.3} becomes clear upon heating the sample quasistatically until the domain almost completely melts into the nematic, and then cooling it marginally by, say, $0.2\,\degree\mathrm{C}$.  Some of the \TBN nuclei still surviving on a submicroscopic scale in the nematic fluid will now begin to grow rapidly.  We illustrate this growth process in Fig.~\ref{fig:Fig.7}.
Initially, as seen in Figs.~\ref{fig:Fig.7}a and \ref{fig:Fig.7}b, the central region of a growing \TBN germ presents a nearly elliptical appearance with the major axis along $x$; spiked finger-like extensions on either side along the minor axis $y$ mark the predominant directions of growth. Although the temperature gradient is radial, notably the growth pattern is not circularly symmetric. In fact, \emph{fingers} growing along $\pm x$ are also observed, but they are slower and develop fast-extending side branches along $\pm y$. This mode of \TBN growth is reminiscent of dendritic morphology, except that side branching, which is conspicuous for fingers growing along $\pm x$, is almost completely suppressed for those growing along $\pm y$. As in dendrites, the growing tips here too are often found to be parabolic. The aforesaid stripe defects develop with the fingers, nearly along $x$ everywhere; they are readily seen forming across the fingers growing along $\pm y$ and drifting in the direction of growth.

Clearly, what we see in Fig.~\ref{fig:Fig.3} is the nonequilibrium shape dominated by anisotropic kinetic factors rather than interfacial energy. The directed growth leading to this geometry occurs along the normal to the nematic director along which the thermal diffusivity is the lowest.\cite{pieranski:static} This \emph{inverted growth}, characterized by the maximum velocity being in the direction of least efficient removal of latent heat, may appear surprising at first sight. However, as pointed out in the context of a similar growth of a smectic B liquid crystal within its own nematic melt,\cite{kurnatowski:selection} global heat removal is maximal when the diffusivity is the largest along the orthogonal to the extended flanks of the growing fingers.

\begin{figure}[ht]
  \centering
  \includegraphics[width=.8\linewidth]{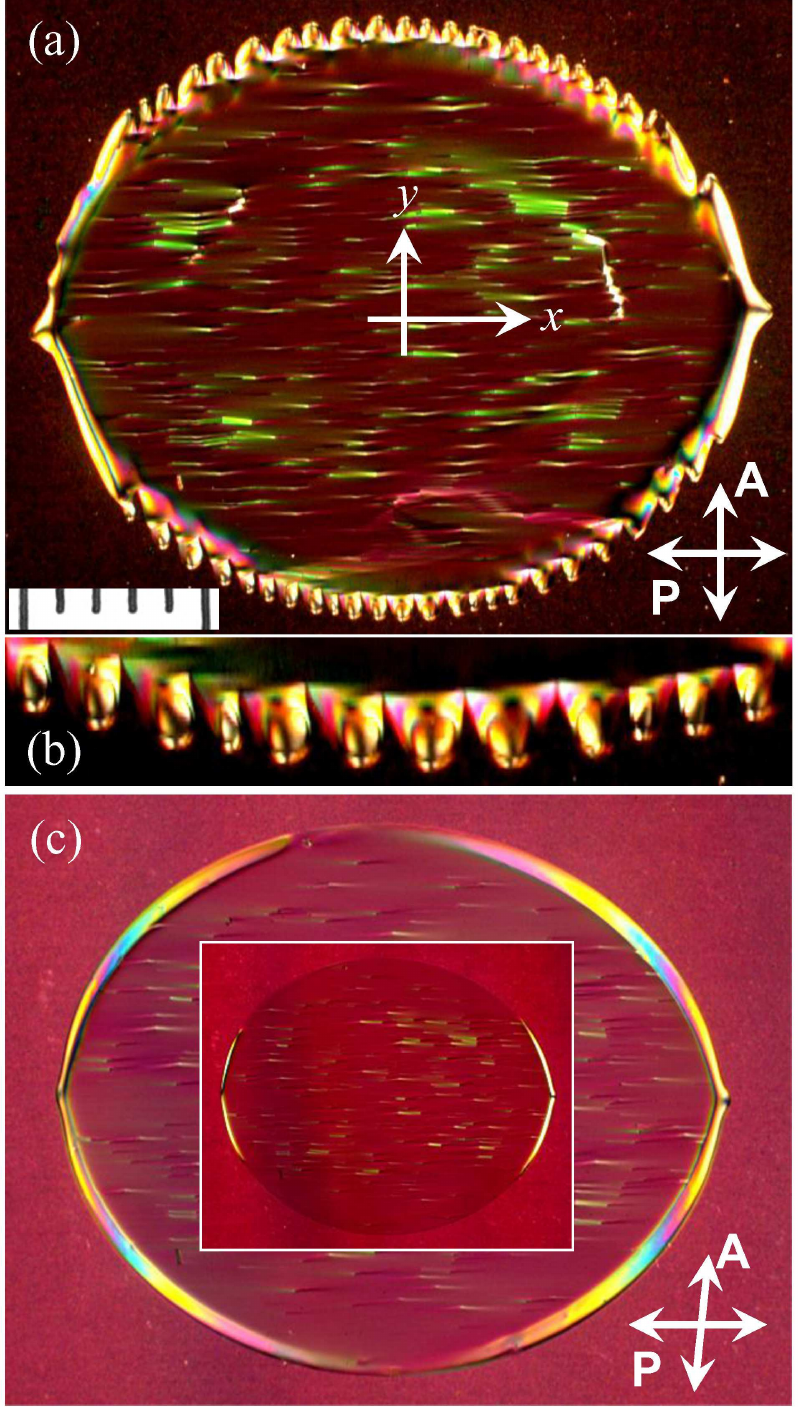}
  \caption{
  (a) \TBN domain surrounded by the nematic melt at about $T=103.1\,\degree\mathrm{C}$.  (b) Lower serrated boundary region (with $3\times$ enlargement) showing dark crosses at the parabolic tips.  (c) \TBN domain at $T=103.2\degree\mathrm{C}$ with two cusps along $x$ and a birefringent band along the interface. Inset: Same as (c) but for the diminished birefringent band. Each scale div. represents $0.1\,\mathrm{mm}$ in (a) and (c) and $0.2\,\mathrm{mm}$ in the inset.}
\label{fig:Fig.8}
\end{figure}

\begin{figure}[ht]
  \centering
  \includegraphics[width=.8\linewidth]{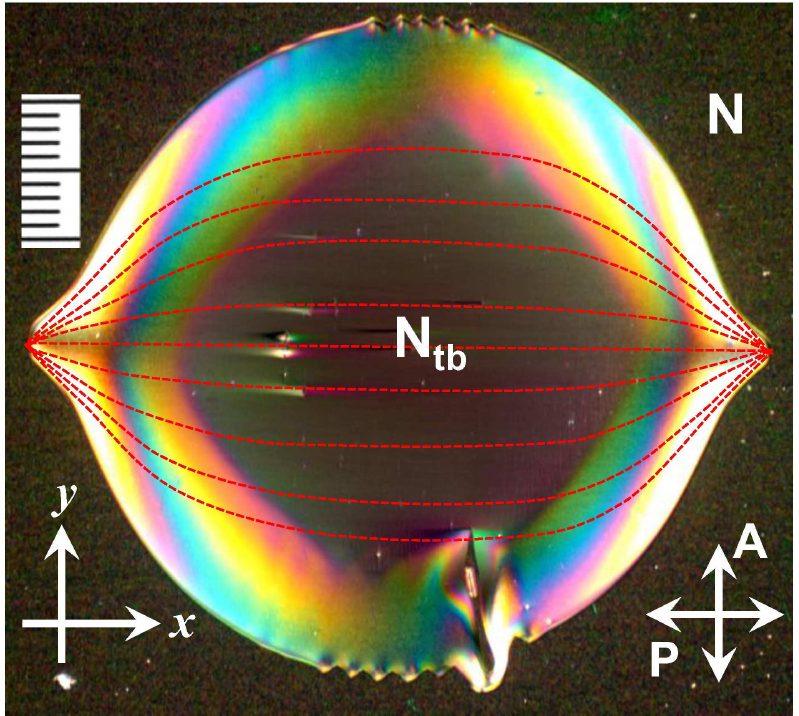}
  \caption{
  \TBN drop almost at temperature \Tm\ showing angular terminations along the easy axis of the surrounding nematic liquid crystal. Dotted lines indicate the possible twist field of the bipolar drop that is nearly free of twin-stripe defects. $20\,\mu\mathrm{m}$ each scale div.}
\label{fig:Fig.9}
\end{figure}

It is possible to realize the near-equilibrium shape of the \TBN domain, which is determined mainly by anisotropic interfacial energy, starting with such a domain as in Fig.~\ref{fig:Fig.7}d. When the latter is very slowly heated to a temperature close to \Tm\ to within $0.2\,\degree\mathrm{C}$, the fingers slowly recede and altogether disappear in a matter of hours. The overall geometry of the drop then displays two cusps at the extremities of the major axis, which happens to be along the rubbing direction. This transformation is exemplified in Fig.~\ref{fig:Fig.8}.
In Fig.~\ref{fig:Fig.8}a, the drop appears serrated at the top and bottom boundaries due to the remnant finger-tips directed along $\pm y$. That these tips are associated with topological defects is borne by the dark crosses they display corresponding to the axes of the crossed polarizers (Fig.~\ref{fig:Fig.8}b). Over a few hours, the boundary of the drop turns smooth except for the two cusps along $x$ (Fig.~\ref{fig:Fig.8}c). A conspicuous feature at this stage is the presence of a birefringent band at the interface that seems to delineate the meniscus. As the drop stabilizes, this band diminishes to a large extent as indicated in the inset to Fig.~\ref{fig:Fig.8}c. Thermal fluctuations often manifest in the momentary reappearance of this band or of the fingers. The density of twin-stripes is another indicator of temperature; it reduces on approaching \Tm. Thus, on elevating the temperature slightly above $T=103.2\,\degree\mathrm{C}$ (e.g., via increased light intensity), while still preserving the \TBN domain, the drop becomes practically defect free in the interior. This situation is exemplified in Fig.~\ref{fig:Fig.9}, wherein the drop has only a few twin stripes inside, with the peripheral zone being completely defect free.
Secondly, in the region of cusps, interference colour rises radially inward and this does not follow from the temperature gradient involved, since the birefringence in the \TBN phase is known to drop with decreasing $T$.\cite{meyer:temperature}  In all probability, what we have in Fig.~\ref{fig:Fig.9} is essentially a bipolar drop having the twist director field as schematically depicted by the dotted lines. This explains the sequence of colours found in the cusp-regions. The equilibrium shape of the \TBN drops illustrated in Figs.~\ref{fig:Fig.8} and \ref{fig:Fig.9} will be described analytically by the mathematical model proposed in the theoretical Sec.~\ref{sec:theory} below.

\begin{figure}[ht]
  \centering
  \includegraphics[width=.8\linewidth]{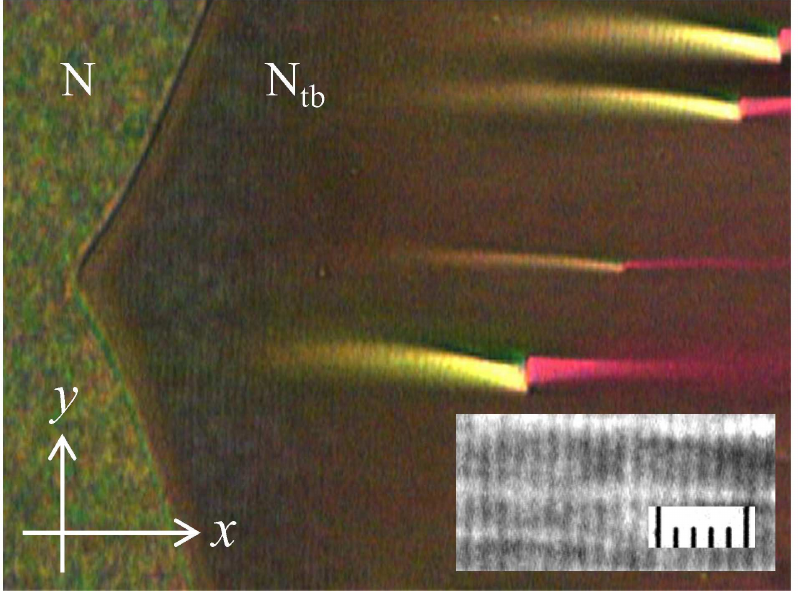}
  \caption{
 The \TBN phase is patterned close to its melting point, showing periodic close-spaced stripes that run perpendicular to the $x$-axis. The inset is an enlargement of the stripes present in the cusp region. $2\,\mu\mathrm{m}$ each scale div.}
\label{fig:Fig.10}
\end{figure}

Another noteworthy morphological feature of the \TBN phase is that it is patterned in the vicinity of \Tm.  As illustrated in Fig.~\ref{fig:Fig.10}, the pattern consists of periodic closely-spaced stripes, with the wave vector along the nematic director.
\begin{figure}[ht]
  \centering
\includegraphics[width=.8\linewidth]{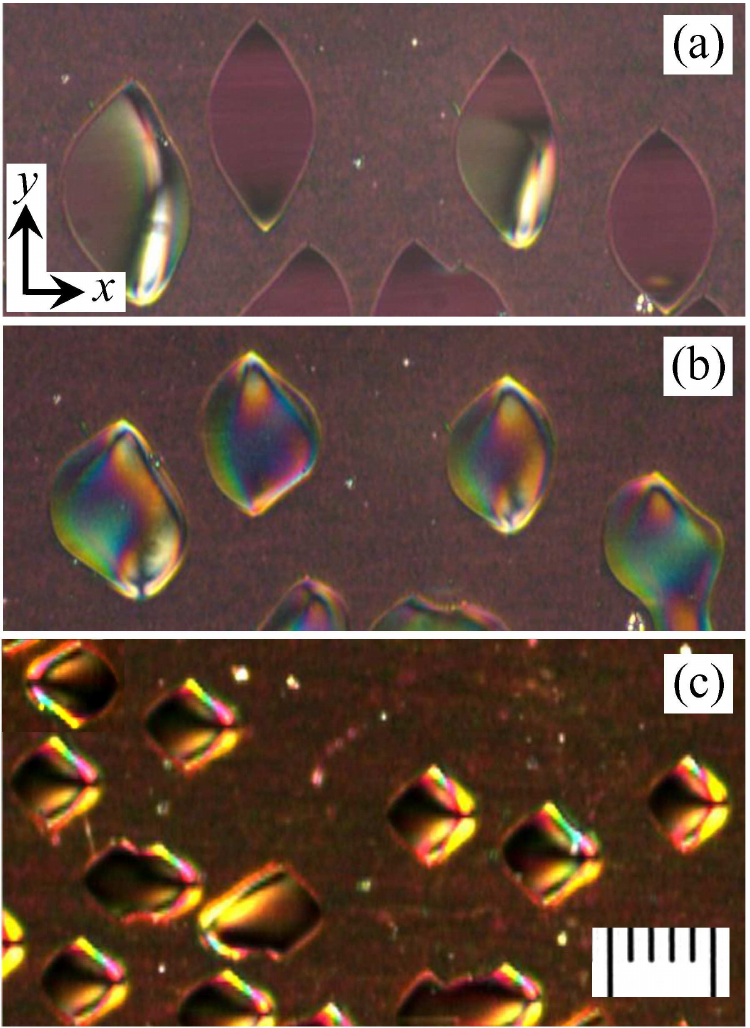}
  \caption{
(a) Disintegration of a mm-size \TBN domain under heating into small, tactoid-shaped drops with vertices along the major axis parallel to $y$. (b) Complex distortions occurring in the drops in (a),  in the process of their melting into nematic. (c) Quadrupole defect structure of the drops of size comparable to the sample thickness $d=20\,\mu\mathrm{m}$. Analyzer along $y$ and polarizer nearly crossed. $10\,\mu\mathrm{m}$ each scale div.}
\label{fig:Fig.11}
\end{figure}
This aspect, also noted earlier,\cite{panov:hierarchy} may indicate the presence of microscale undulations that disappear on further cooling. It is unlikely to bear any relation to the nanoscale helicity of the phase.

When a mm-size \TBN drop of the type in Fig.~\ref{fig:Fig.9} at a temperature close to melting is exposed to further enhanced light intensity, under the mild heating caused thereby, it disintegrates into micron-sized droplets as in Fig.~\ref{fig:Fig.11}a.
Interestingly, the droplets, which are again with two cusps, have their vertices along the major axis now parallel to $y$. As the droplets melt, they undergo complex  distortions manifesting in  shifting birefringence colours and defects exemplified in Fig.~\ref{fig:Fig.11}b. Very small drops (Fig.~\ref{fig:Fig.11}c) are about the same size along $x$ and $y$, and appear to involve four vertices. In passing, we may mention that nematic drops surrounded by the \TBN phase are also tactoid-shaped, similar to the \TBN drops in Fig.~\ref{fig:Fig.11}a.

\begin{figure}[ht]
  \centering
  \includegraphics[width=.8\linewidth]{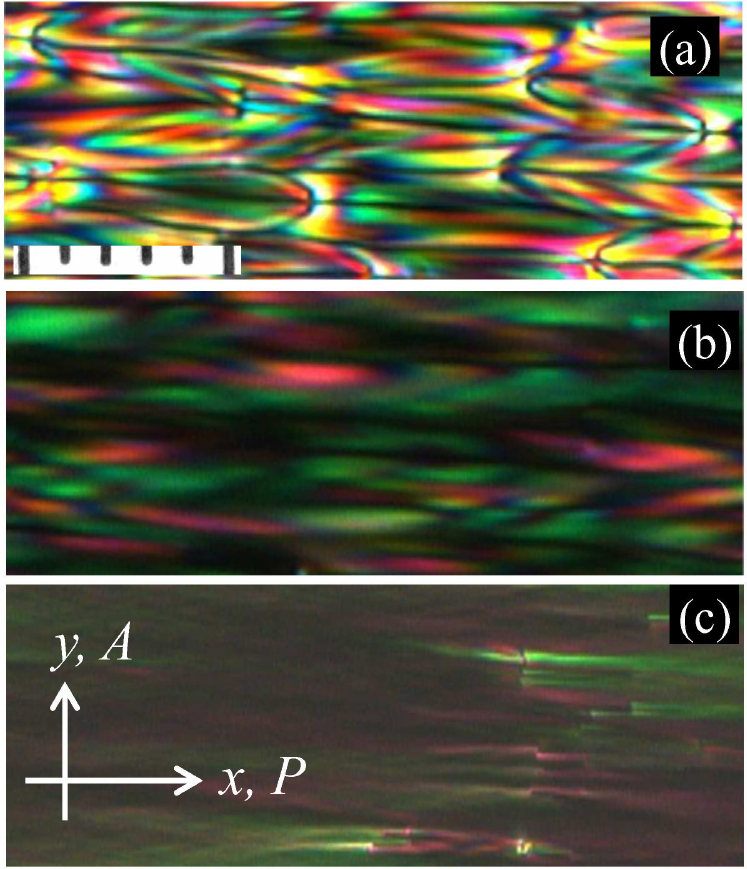}
  \caption{
(a) Focal conic texture of the \TBN phase formed at $T=102.5\,\degree\mathrm{C}$. (c) The twin-stripe texture into which the texture in (a) transforms upon increasing $T$ by $\sim0.6\,\degree\mathrm{C}$. (b) Intermediate texture observed in the course of changing $T$. $5\,\mu\mathrm{m}$ each scale div.}
\label{fig:Fig.12}
\end{figure}

Lastly, it seems significant to mention a temperature determined morphological transformation within the \TBN phase, although it is not linked to the interfacial geometry of \TBN drops. We refer here to the twin-stripe texture such as seen very close to \Tm\ (see, for example, Fig.~\ref{fig:Fig.6}a).  When $T$ is reduced to about $0.5\,\degree\mathrm{C}$ below \Ts, the morphology changes strikingly in that the twin-stripes transform into familiar focal conic defects.\cite{challa:twist-bend} This change is reversible as demonstrated in Fig.~\ref{fig:Fig.12}.
The structural implication of this morphological switch remains to be understood.

\section{Theory}\label{sec:theory}
The twist director $\twist$ is likely to designate the optic axis of \TBN phases, as the heliconical arrangement of the molecular director $\mol$ and its possible elastic distortions are associated with a typical length scale much shorter than optical wavelengths. Here, before solving analytically a 2D mathematical model for some of the tactoidal equilibrium shapes described above, we pause to derive an appropriate interfacial energy density featuring $\twist$ and the outward unit normal $\normal$ to the drop.

\subsection{Surface energy functional}\label{sec:surface_energy_functional}
The surface separating a \TBN drop from a nematic environment bears an interfacial energy whose density per unit area $\Ws$ can be written in the classical Rapini-Papoular form
\begin{equation}\label{eq:surface_energy_density}
\Ws=W_0\left[1+w(\mol\cdot\normal)^2\right],
\end{equation}
where $W_0$ is a the isotropic surface tension, $\mol$ is the molecular nematic director, $\normal$ is the outer unit normal to the drop's boundary, and $w\geqq-1$ is a dimensionless parameter weighting the anisotropic component of the surface tension against the isotropic one. In particular, for $w>0$, the surface would induce a degenerate tangential alignment of $\mol$, whereas for $w<0$ it would induce the homeotropic alignment.

While $\mol$ should be considered to be continuous across the $\mathrm{N_{tb}}$/N interface and to agree with the uniform nematic director $\n=\ave{\mol}$ outside the drop, inside the drop it is expected to comply, at least locally, with the heliconical ground state of twist-bend nematics described by
\begin{equation}\label{eq:NTB_gorund_state}
\mol=\cos\ca\,\twist+\sin\ca\left(\cos\az\,\e_1+\cos\az\,\e_2\right),
\end{equation}
where $\twist$ is the \emph{twist director} designating the optic axis of the \TBN phase, $\frame$ is a frame orthogonal to $\twist$, $\ca$ is the \emph{cone angle} characteristic of the phase, and $\az$ is a field varying in space over a nanoscopic length-scale. At such a length scale, \eqref{eq:surface_energy_density} can be viewed as a field in the vicinity of the interface, with $\normal$ uniform in space. Coarse-graining $\Ws$ would then simply amount to averaging $\az$, yielding
\begin{equation}\label{eq:coarse-grained_Ws}
\ave{\Ws}=W_0\left\{1+\frac12w\sin^2\ca-\frac12w\left(1-3\cos^2\ca\right)(\twist\cdot\normal)^2 \right\}.
\end{equation}
Assuming $w>0$ and $\cos\ca>\frac{1}{\sqrt{3}}$, $\ave{\Ws}$ is minimized by letting $\twist$ be tangent to the interface, however oriented upon it. As suggested by the experimental observations, in the following we shall assume that $\twist$ is subject to such a \emph{degenerate} tangential anchoring on the drop's free surface, which the coarse-grained energy in \eqref{eq:coarse-grained_Ws} also contribute to justify. In view of the uniform alignment of the nematic director $\n$ outside the drop and the continuity of the molecular alignment across the interface, the degenerate tangential anchoring of $\twist$ becomes geometrically compatible only if the cone angle is allowed to vary along the interface, at the expenses of an elastic \emph{mismatch} energy. To the surface density $f_\mathrm{a}$ of such an energy we shall give the simplest quartic form,
\begin{equation}\label{eq:mismatch_energy_density}
f_\mathrm{a}=\frac12\Wa\left[(\twist\cdot\n)^2-c^2\right]^2,
\end{equation}
where $\Wa>0$ is an elastic constant that penalizes the departure of $\twist$ from making the ideal cone angle $\ca=\arccos c$ with the nematic director $\n$. Compatibility with \eqref{eq:coarse-grained_Ws} requires that $c$ satisfy $c>\frac{1}{\sqrt{3}}$.

In summary, the surface energy functional that we consider here for the interface separating a \TBN drop from a nematic environment fully aligned along $\n$ is
\begin{equation}\label{eq:interface_functional}
\ens=W_0\int_\surface\left\{1+\frac12\omega\left[(\twist\cdot\n)^2-c^2\right]^2\right\}da,
\end{equation}
where we have set
\begin{equation}\label{eq:omega_definition}
\omega:=\frac{\Wa}{W_0}>0,
\end{equation}
with $\Wa$ as in \eqref{eq:mismatch_energy_density} and $W_0$ an isotropic surface tension,\footnote[3]{By \eqref{eq:coarse-grained_Ws}, $W_0$ in \eqref{eq:omega_definition} has properly been rescaled by the factor $1+\frac12w\sin^2\ca$ compared to $W_0$ in \eqref{eq:surface_energy_density}.} $\surface$ is the surface separating the two phases and $a$ is the area measure over it. The functional $\ens$ in \eqref{eq:interface_functional} represents the surface energy of the drop where the classical isotropic surface tension is supplemented by the energy needed to distort locally the equilibrium heliconical ground state of the \TBN phase, under the assumption that the coarse-grained surface energy imposes the degenerate tangential anchoring for the twist director $\twist$. As the \TBN phase, like the classical nematic phase, can be regarded as incompressible, $\ens$ is subject to the constraint that the volume enclosed by $\surface$ be fixed.

Clearly, the $\twist$ field will also be distorted in the interior of the drop. Optical observations of the inner texture of $\twist$ suggest that it might be arranged in a bipolar fashion with poles coincident with the tactoid tips. We shall hereafter neglect the elastic energy associated with such a distortion, as we believe that it plays a little role in determining the equilibrium shape of a \TBN drop, which is mainly driven by the interfacial energy. Thus, hereafter $\ens$ will remain the only energy functional for our equilibrium problem.

\subsection{Two-dimensional drops}\label{sec:two_dimensional_drops}
The functional in \eqref{eq:interface_functional} has a natural, dimensionless version which is appropriate for the 2D drops described in this paper. It takes the following form
\begin{equation}\label{eq:F_functional}
F[\rv]:=\int_0^L\left\{1+\frac12\omega\left[(\twist\cdot\n)^2-c^2\right]^2\right\}ds,
\end{equation}
where $L$ is the (undetermined) length of the curve $\curve$ bounding the drop, $\twist$ now coincides with a unit tangent vector to $\curve$, and $s$ is the corresponding arc-length co-ordinate, so that $\curve$ is described by the mapping $s\mapsto\rv(s)$ and $\twist=\rv'$, where a prime $'$ denotes differentiation with respect to $s$.

\begin{figure}[ht]
  \centering
  \subfigure[]{\includegraphics[width=.7\linewidth]{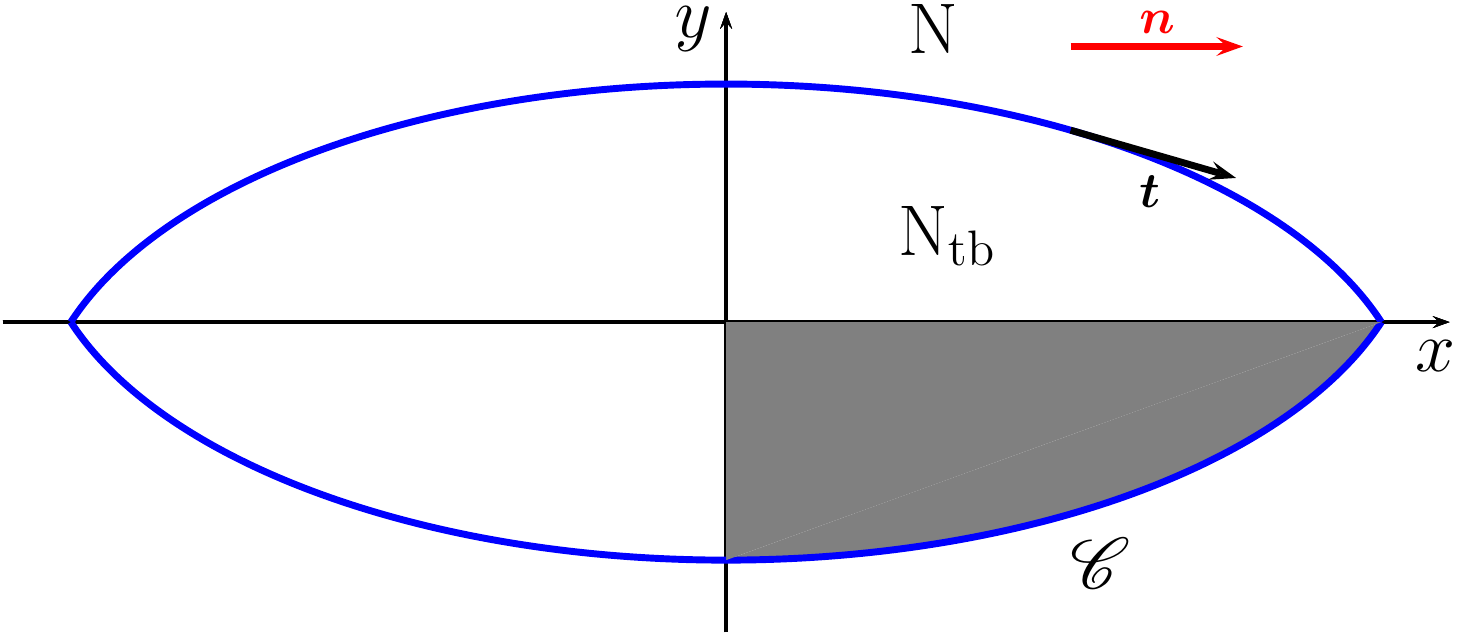}}\\
  \hspace{.05\linewidth}
  \subfigure[]{\includegraphics[width=.4\linewidth]{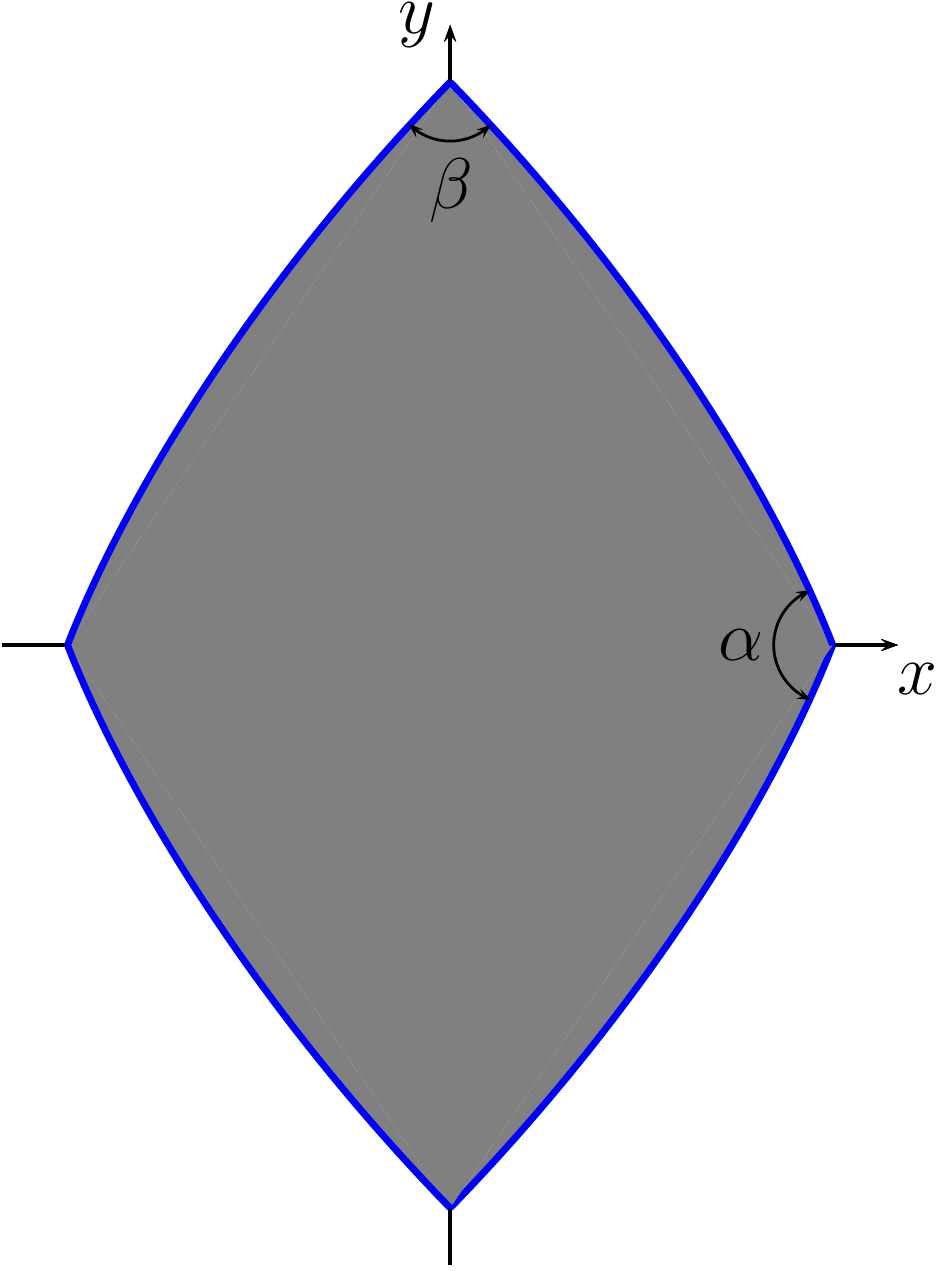}}
  \caption{
  (a) An admissible curve $\curve$ with a two-fold symmetry. The symmetry axis $x$ designates the orientation of the uniformly aligned nematic phase, N. The region delimited by $\curve$ is occupied by the \TBN phase; $\twist$ is the tangent unit vector to $\curve$, to be identified with the twist director on the drop's interface. The specific curve $\curve$ depicted here has a corner where it meets the $x$ axis, whereas it is smooth where it meets the $y$ axis. The gray quarter, which repeated reflections make cover the whole drop, admits a Cartesian representation, $y=y(x)$. (b) A two-fold symmetric shape exhibiting four corners on the symmetry axes; $\alpha$ and $\beta$ are the corresponding inner corner angles.}
\label{fig:quarter_drop}
\end{figure}

Figure~\ref{fig:quarter_drop}(a) illustrates an admissible shape for $\curve$, symmetric with respect to both
axes $x$ and $y$, the former designating the orientation of $\n$. The area $A$ enclosed by $\curve$ can be expressed in terms of $\rv$ as
\begin{equation}\label{eq:area_functional}
A[\rv]:=-\frac12\int_0^L\rv\times\twist\cdot\ez ds,
\end{equation}
where $\ez:=\ex\times\ey$.

Constrained equilibrium for $F$ requires that $\rv$ make the first variations $\delta F$ and $\delta A$ proportional to one another,
\begin{equation}\label{eq:variations_proportionality}
\delta F=\lambda\delta A,
\end{equation}
where $\lambda$ is a Lagrange multiplier, still to be determined. The first variation $\delta F$ is a functional, $\delta F(\rv)[\uv]$, linear in the variation $\uv$ of $\rv$. Formally,
\begin{equation}\label{eq:delta_F_definition}
\delta F(\rv)[\uv]:=\left.\frac{d}{ds}F[\rv_\vae]\right|_{\vae=0},
\end{equation}
where $\rv_\vae:=\rv+\vae\uv$ and $\vae$ is a small, perturbation parameter. The perturbed curve $\curve_\vae$ described by $\rv_\vae$ has unit tangent vector $\twist_\vae$ delivered by
\begin{equation}\label{eq:t_epsilon}
\twist_\vae=\twist+\vae(\Pt)\uv'+o(\vae),
\end{equation}
where $\Id$ is the identity tensor. Moreover, the local dilation ratio between lengths along $\curve_\vae$ and lengths along $\curve$ is, to within first order in $\vae$, $1+\vae\twist\cdot\uv'$. Thus, by letting $\n=\ex$ from \eqref{eq:F_functional} and \eqref{eq:delta_F_definition} we obtain
\begin{equation}\label{eq:delta_F_computation}
\begin{split}
\delta F(\rv)[\uv]=\int_0^L\left\{2\omega\left[(\twist\cdot\ex)^2-c^2\right](\Pt)\ex\right. \\ +\left.\left(1+\frac12\omega\left[(\twist\cdot\ex)^2-c^2\right]^2\right)\twist\right\}\cdot\uv'ds.
\end{split}
\end{equation}
Similarly, we arrive at
\begin{equation}\label{eq:delta_A_computation}
\delta A(\rv)[\uv]=\frac12\int_0^L\left(\normal\cdot\uv-\ez\times\rv\cdot\uv'\right)ds,
\end{equation}
where $\normal:=\ez\times\twist$ is the outer unit normal to $\curve$.

As shown in Fig.~\ref{fig:quarter_drop}(a), $\curve$ may possess a \emph{corner}, that is a point, say at $s=s_0$, where the unit tangent $\twist$ jumps from $\tm$ to $\tp$ as $s$ increases through $s_0$. If this is the case, splitting the integral in  \eqref{eq:delta_A_computation} into subintervals where $\twist$ is continuous, allowing both $\uv$ and $\uv'$ to be everywhere continuous, and integrating by parts, we easily show that
\begin{equation}\label{eq:delta_A_computation_final}
\delta A(\rv)[\uv]=\int_0^L\normal\cdot\uv ds.
\end{equation}
Contrariwise, proceeding just in the same way, we extract a \emph{jump} contribution to $\delta F(\rv)$ from every point of discontinuity for $\twist$, which reads as $\jump{\f}\cdot\uv$, where
\begin{equation}\label{eq:f_definition}
\begin{split}
\f:=2\omega[(\twist\cdot\ex)^2-c^2](\twist\cdot\ex)(\Pt)\ex \\ +\left(1+\frac12\omega[(\twist\cdot\ex)^2-c^2]^2\right)\twist
\end{split}
\end{equation}
and, as customary, for any discontinuous field $\psi$ the \emph{jump} $\jump{\psi}$ is defined by $\jump{\psi}:=\psi^+-\psi^-$, where $\psi^+$ and $\psi^-$ are the right and left limits of $\psi$ at the point of discontinuity.

Requiring \eqref{eq:variations_proportionality} to be valid for arbitrary $\uv$, by \eqref{eq:delta_F_computation} and \eqref{eq:delta_A_computation_final} we conclude that the equilibrium equation for the regular arcs of $\curve$, where $\twist$ is continuous, is
\begin{equation}\label{eq:equilbrium_regular_arcs}
\f'+\lambda\normal=\bm{0},
\end{equation}
while the equation
\begin{equation}\label{eq:jump_f_0}
\jump{\f}=\bm{0}
\end{equation}
must hold at all corners, where $\twist$ is discontinuous. Both equations \eqref{eq:equilbrium_regular_arcs} and \eqref{eq:jump_f_0} suggest to interpret $\f$ as a \emph{line stress}. As shown below, equations \eqref{eq:equilbrium_regular_arcs} and \eqref{eq:jump_f_0} can be solved exactly. If \eqref{eq:equilbrium_regular_arcs} is essentially equivalent to Wulff's construction in two space dimensions, for which an analytic characterization has long been known,\cite{burton:growth,kim:morphogenesis} the jump condition \eqref{eq:jump_f_0} is more in the spirit of the mechanics of discontinuities, as recently revived in a number of studies, some also entailing striking experimental consequences.\cite{biggins:understanding,biggins:growth,virga:dissipative,virga:chain,hanna:jump}

\subsection{Equilibrium corners}\label{sec:equilibrium_corners}
In general, the equilibrium equations \eqref{eq:equilbrium_regular_arcs} and \eqref{eq:jump_f_0} are rather complicated. Symmetry may simplify them. In keeping with the experimental observations, we shall hereafter assume that $\curve$ has the two-fold symmetry displayed in Fig.~\ref{fig:quarter_drop} and that its corners may only occur on the symmetry axes.

For a corner on the $y$ axis, $\tm$ and $\tp$ satisfy
\begin{equation}\label{eq:y_corner}
\tp\cdot\ex=\tm\cdot\ex,\quad\tp\cdot\ey=-\tm\cdot\ey,
\end{equation}
and it can be shown that equation \eqref{eq:jump_f_0} reduces to
\begin{equation}\label{eq:chi_y_corner}
3\omega\chi^2-2\omega c^2\chi-2-\omega c^4=0,
\end{equation}
where we have set $\chi:=(\twist\cdot\ex)^2$. It can be easily proved that there is precisely one root of \eqref{eq:chi_y_corner} in $[0,1]$, which reads
\begin{equation}\label{eq:chi_1}
\chi=\chi_1(c,\omega):=\frac13\left(c^2+\sqrt{4c^4+\frac6\omega}\right),
\end{equation}
if and only if
\begin{equation}\label{eq:omega_1}
\omega\geqq\oone(c):=\frac{2}{(1-c^2)(3+c^2)}.
\end{equation}
Otherwise, there is none and at equilibrium no corner can arise where $\curve$ meets the $y$ axis. The inner corner angle $\beta$ depicted in Fig.~\ref{fig:quarter_drop}b is given by
\begin{equation}\label{eq:beta_formula}
\beta=2\arcsin\sqrt{\chi_1},
\end{equation}
where $\chi_1$ is as in \eqref{eq:chi_1}; in particular, \eqref{eq:beta_formula} implies that
\begin{equation}\label{eq:beta_limit}
\lim_{\omega\to\infty}\beta(c,\omega)=2\arcsin c.
\end{equation}

Similarly, for a corner on the $x$ axis,
\begin{equation}\label{eq:x_corner}
\tp\cdot\ex=-\tm\cdot\ex,\quad\tp\cdot\ey=\tm\cdot\ey
\end{equation}
and \eqref{eq:jump_f_0} reduces to
\begin{equation}\label{eq:chi_x_corner}
3\omega\chi^2-2\omega(c^2+2)\chi-(\omega c^4-4\omega c^2+2)=0.
\end{equation}
It can be easily proved that there is precisely one admissible root of \eqref{eq:chi_x_corner}, which reads
\begin{equation}\label{eq:chi_2}
\chi=\chi_2(c,\omega):=\frac13\left(c^2+2-\sqrt{4(1-c^2)^2+\frac6\omega}\right),
\end{equation}
if and only if
\begin{equation}\label{eq:omega_2}
\omega\geqq\otwo(c):=\frac{2}{c^2(4-c^2)}.
\end{equation}
Otherwise, there is none and at equilibrium no corner can arise where $\curve$ meets the $x$ axis. The inner corner angle $\alpha$ depicted in Fig.~\ref{fig:quarter_drop}b is given by
\begin{equation}\label{eq:alpha_formula}
\alpha=2\arccos\sqrt{\chi_2},
\end{equation}
where $\chi_2$ is as in \eqref{eq:chi_2}; in particular, \eqref{eq:alpha_formula} implies that
\begin{equation}\label{eq:alpha_limit}
\lim_{\omega\to\infty}\alpha(c,\omega)=2\arccos c=2\ca,
\end{equation}
where $\ca$ is the ideal cone angle.

\begin{figure}[ht]
  \centering
  \includegraphics[width=\linewidth]{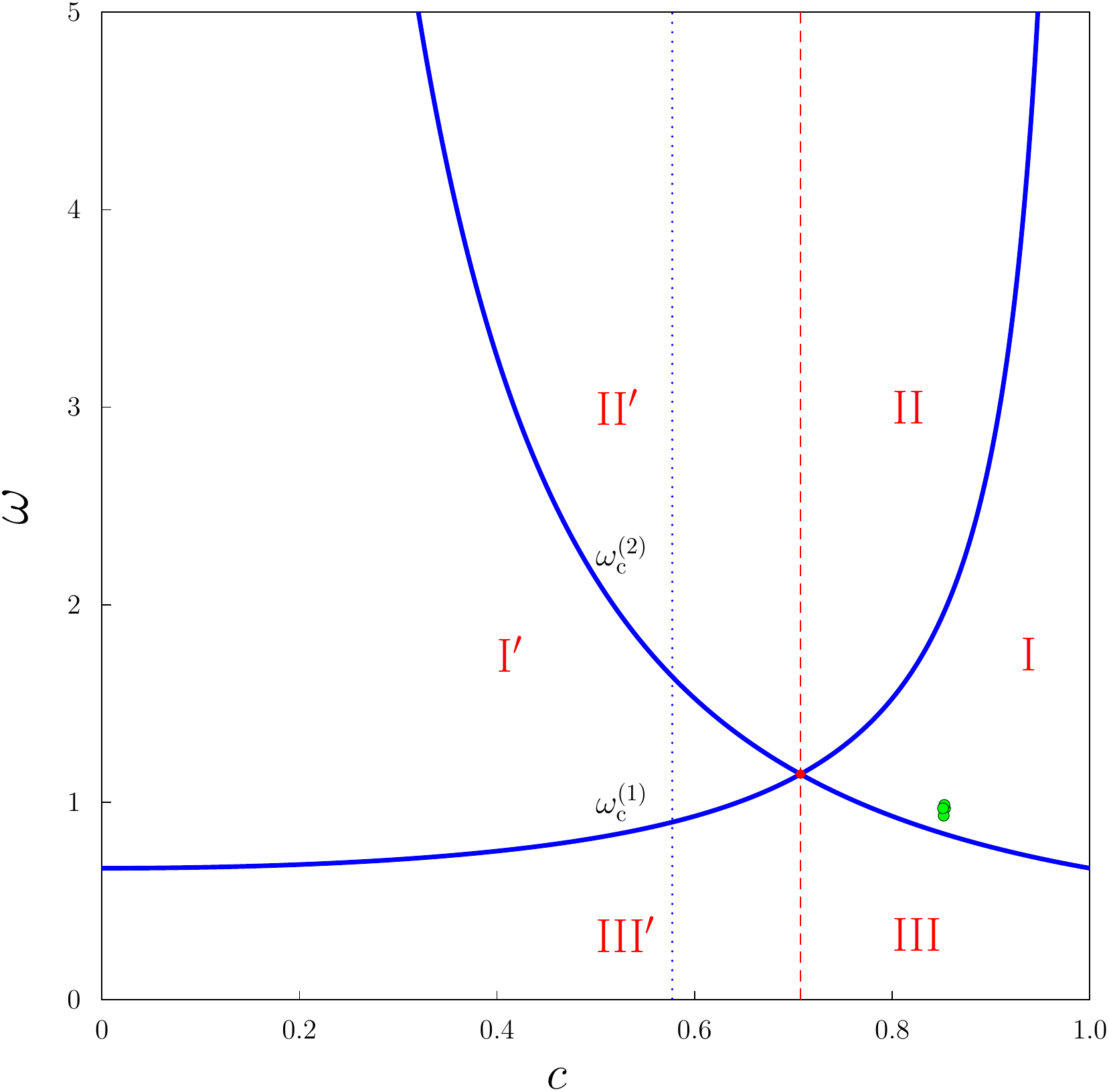}
  \caption{
  Phase diagram in the plane $(c,\omega)$. Regions I, II, and III ranging over $\frac{1}{\sqrt{2}}\leqq c\leqq1$ are delimited by the graphs of the functions $\oone$ and $\otwo$ in \eqref{eq:omega_1} and \eqref{eq:omega_2}. These graphs, which cross at the (red) triple point $(\frac{1}{\sqrt{2}},\frac87)$, are transformed one into the other by the mapping $c\mapsto\sqc$. The corresponding images I$'$, II$'$, and III$'$ of regions I, II, and III, on the other side of the dashed separatrix $c=\frac{1}{\sqrt{2}}$, cover together with I, II, and III the whole admissible parameter plane $(c,\omega)$. The dotted line is drawn at $c=\frac{1}{\sqrt{3}}$: the half-plane $c>\frac{1}{\sqrt{3}}$ complies with the request that the coarse-grained surface energy in \eqref{eq:coarse-grained_Ws} be minimum. The equilibrium shape of the drop has no corners in III and two and four symmetric corners in I and II, respectively. The (green) circles in region I correspond to the following data for the pair of measures $(\rho,\alpha)$: $(1.212,151.5\degree)$, $(1.203,156\degree)$, $(1.214,150\degree)$, $(1.208,152\degree)$.}
\label{fig:phase_diagram}
\end{figure}

\begin{figure}[ht]
  \centering
  \subfigure[]{\includegraphics[width=.8\linewidth]{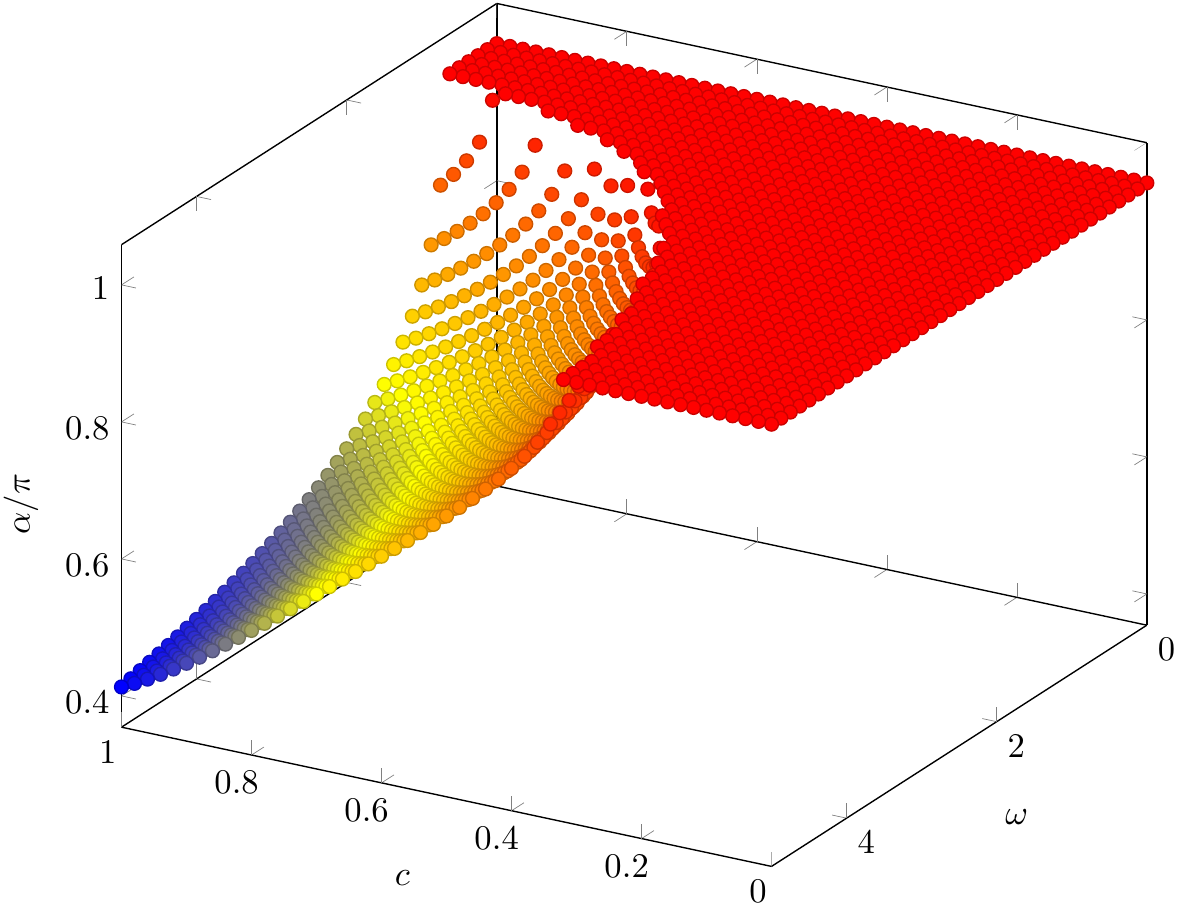}}
  \hspace{.05\linewidth}
  \subfigure[]{\includegraphics[width=.8\linewidth]{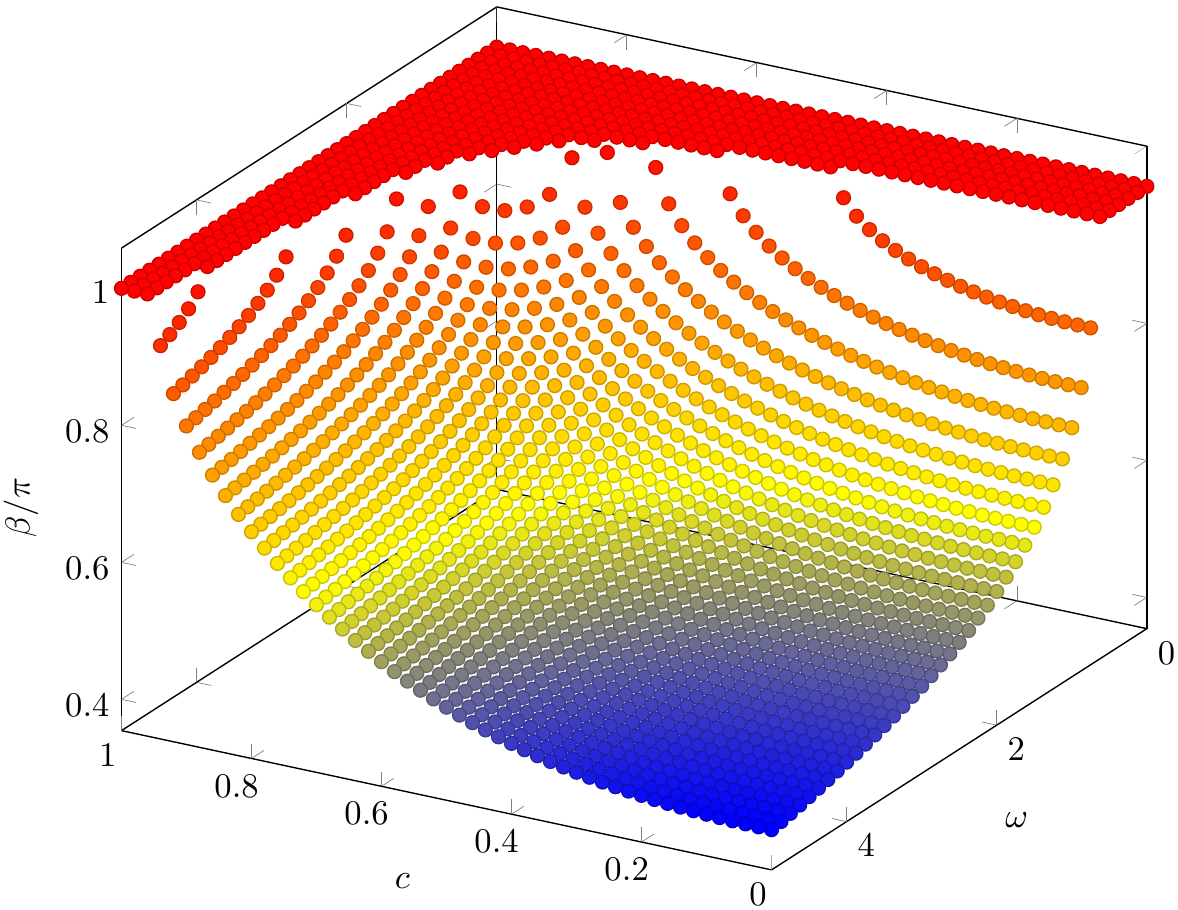}}
  \caption{
  (a) The plot of the corner angle $\alpha$ scaled to $\pi$ as delivered by \eqref{eq:alpha_formula} and \eqref{eq:chi_1}. For $\omega\leqq\otwo$, that is, in region III of Fig.~\ref{fig:phase_diagram}, $\alpha\equiv\pi$ as no equilibrium corner is encountered on the $x$ axis.
  (b) The plot of the corner angle $\beta$ scaled to $\pi$ as delivered by \eqref{eq:beta_formula} and \eqref{eq:chi_2}. For $\omega\leqq\oone$, that is, in regions I and II of Fig.~\ref{fig:phase_diagram}, $\beta\equiv\pi$ as no equilibrium corner is encountered on the $y$ axis.
  }
\label{fig:alpha_beta_plots}
\end{figure}

The graphs of both $\oone$ and $\otwo$ as functions of $c$ are plotted in Fig.~\ref{fig:phase_diagram}.
By direct inspection, we easily realize that $\oone$ and $\otwo$ enjoy the following symmetry property:
\begin{equation}\label{eq:omega_symmetry}
\otwo(c)=\oone\left(\sqc\right),
\end{equation}
so that regions I, II, and III over $\sepa\leqq c\leqq1$ in Fig.~\ref{fig:phase_diagram} together with their images I$'$, II$'$, and III$'$ under the mapping $c\mapsto\sqc$ cover the whole admissible parameter plane $(c,\omega)$. While region III in Fig.~\ref{fig:phase_diagram} represents the locus for the parameters $(c,\omega)$ where no equilibrium corners are possible (on neither of the axes), in region I only corners on the $x$ axis are possible at equilibrium, whereas in region II they are possible on both axes. Moreover, as also suggested by the limits in \eqref{eq:beta_limit} and \eqref{eq:alpha_limit}, the functions $\beta$ and $\alpha$ in \eqref{eq:beta_formula} and \eqref{eq:alpha_formula} obey the property:
\begin{equation}\label{eq:alpha_beta_symmetry}
\beta(c,\omega)=\alpha\left(\sqc,\omega\right),
\end{equation}
for all $\omega\geqq0$. This property is illustrated by the graphs of the functions $\alpha$ and $\beta$ plotted in Fig.~\ref{fig:alpha_beta_plots}.

\subsection{Equilibrium shapes}\label{sec:equilibrium_shapes}
In general, as confirmed by the full analysis presented in the Appendix, in view of the two-fold symmetry assumed for $\curve$ relative to the $x$ and $y$ axes, mapping $c$ into $\sqc$ simply rotates an equilibrium shape of the drop  by $\frac\pi2$ about the $z$ axis. Consequently, finding the equilibrium shapes of $\curve$ for $\sepa\leqq c\leqq1$ amounts to find them all. For this reason, our study can be confined to regions I, II, and III of the parameter plane $(c,\omega)$ in Fig.~\ref{fig:phase_diagram}.

\begin{figure}[ht]
  \centering
  \subfigure[]{\includegraphics[width=.8\linewidth]{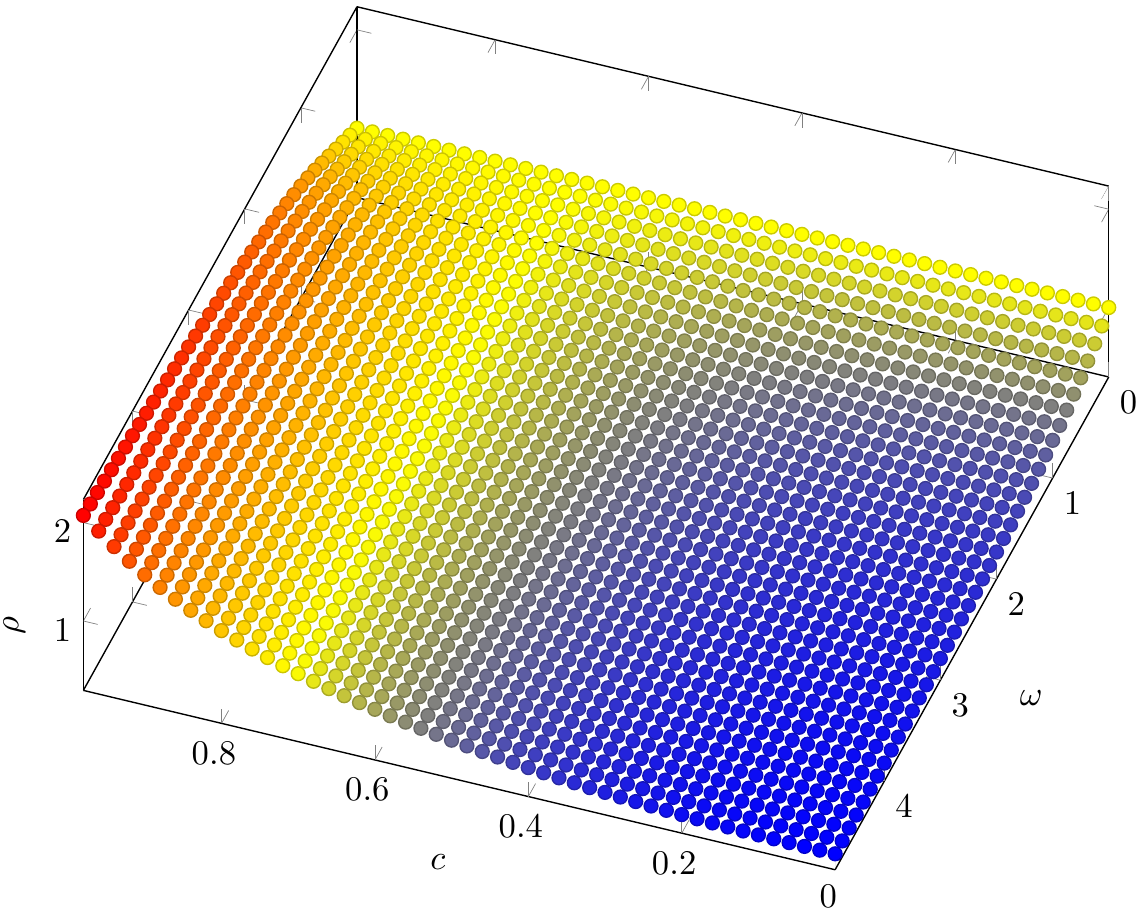}}
  \hspace{.05\linewidth}
  \subfigure[]{\includegraphics[width=.8\linewidth]{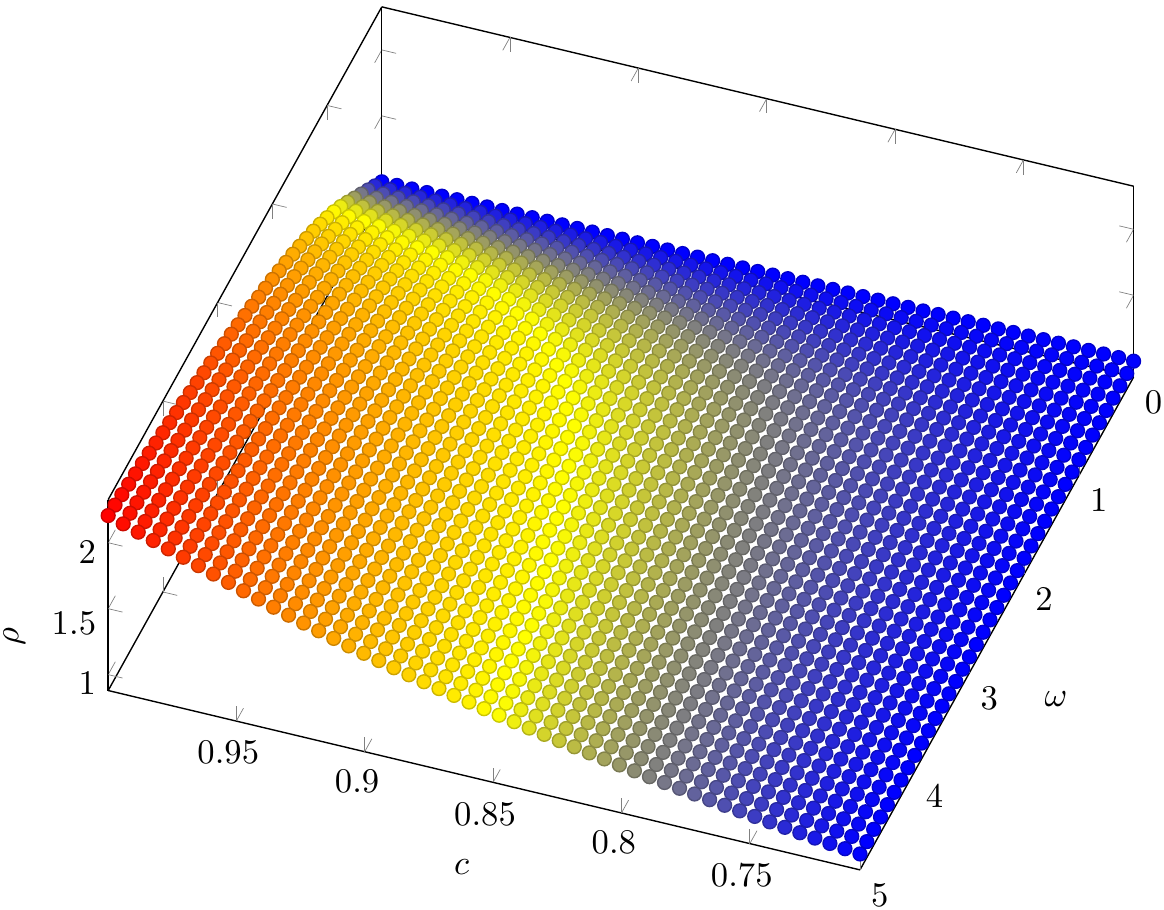}}
  \caption{
  (a) The plot of $\rho$ in the whole admissible parameter plane $(c,\omega)$.
  (b) The plot of $\rho$ in the restricted parameter strip $\frac{1}{\sqrt{2}}\leqq c\leqq1$. By \eqref{eq:rho_symmetry}, this latter suffices to obtain the whole plot in (a); it shows more clearly that $\rho$ satisfies \eqref{eq:rho_separatrix}.
  }
\label{fig:rho}
\end{figure}

A qualitative property of the equilibrium shape of a \TBN drop, which along with the measures of the angles $\alpha$ and $\beta$ delivered by \eqref{eq:alpha_formula} and \eqref{eq:beta_formula} can be used to put our theoretical model to the test is the \emph{aspect ratio} $\rho$ between the extension of $\curve$ along $\n$ and its width across $\n$. As shown in the Appendix, $\rho$ as a function of $(c,\rho)$ enjoys the following symmetry property
\begin{equation}\label{eq:rho_symmetry}
\rho\left(\sqc,\omega\right)=\frac{1}{\rho(c,\omega)},
\end{equation}
which, in particular, entails
\begin{equation}\label{eq:rho_separatrix}
\rho\left(\frac{1}{\sqrt{2}},\omega\right)\equiv1.
\end{equation}
The graph of $\rho$ as obtained from \eqref{eq:rho_formula} and \eqref{eq:u_min_u_max} is plotted in Fig.~\ref{fig:rho}.

By measuring the aspect ratio $\rho$ and the inner corner angle $\alpha$ of the observed tactoids, with the aid of the functions plotted in Figs.~\ref{fig:alpha_beta_plots} and \ref{fig:rho}, we are able to determine the parameters $c=\cos\ca$ and $\omega$ introduced by our model. The almost superimposed circles in Fig.~\ref{fig:phase_diagram} mark the roots $(c,\omega)$ of equations \eqref{eq:rho_formula} and \eqref{eq:alpha_formula} delivering the analytic expressions for $\rho$ and $\alpha$, respectively, for the experimental data $(\rho,\alpha)$ extracted from the images in Fig.~\ref{fig:Fig.8}c (including its inset) and two more not shown here. The average values for the model parameters thus measured are
\begin{equation}\label{eq:measured_parameters}
\ca\approx31.5\degree\quad\text{and}\quad\omega\approx0.96,
\end{equation}
the former agreeing with other, more direct measurements for the same material.\cite{meyer:temperature}

\begin{figure}[ht]
  \centering
  \subfigure[]{\includegraphics[width=.25\linewidth]{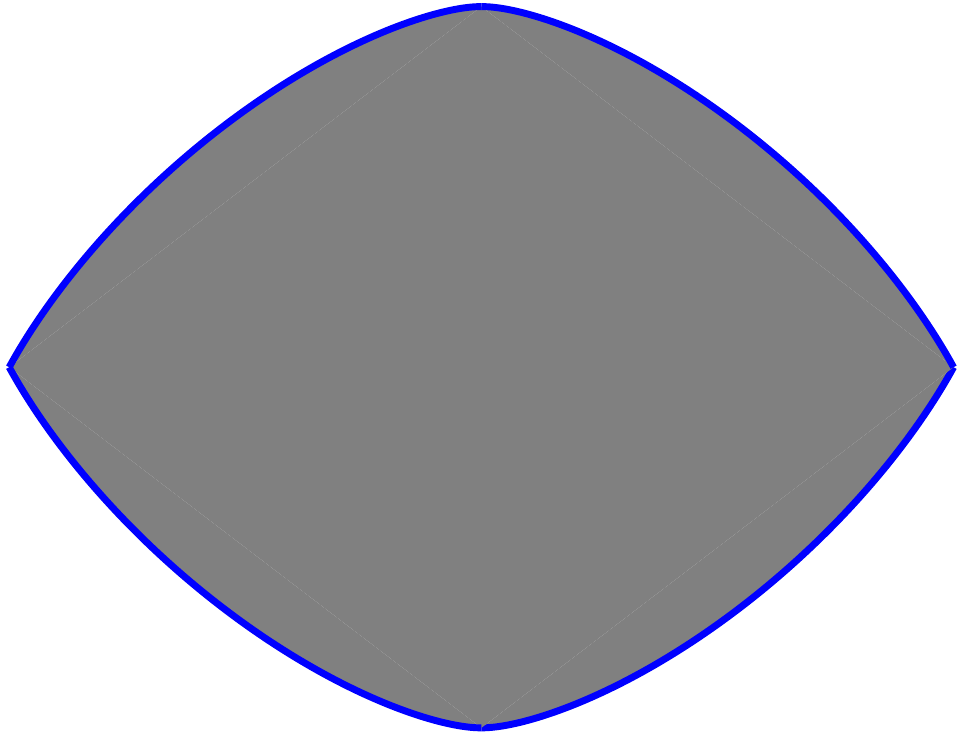}}
  \hspace{.05\linewidth}
  \subfigure[]{\includegraphics[width=.25\linewidth]{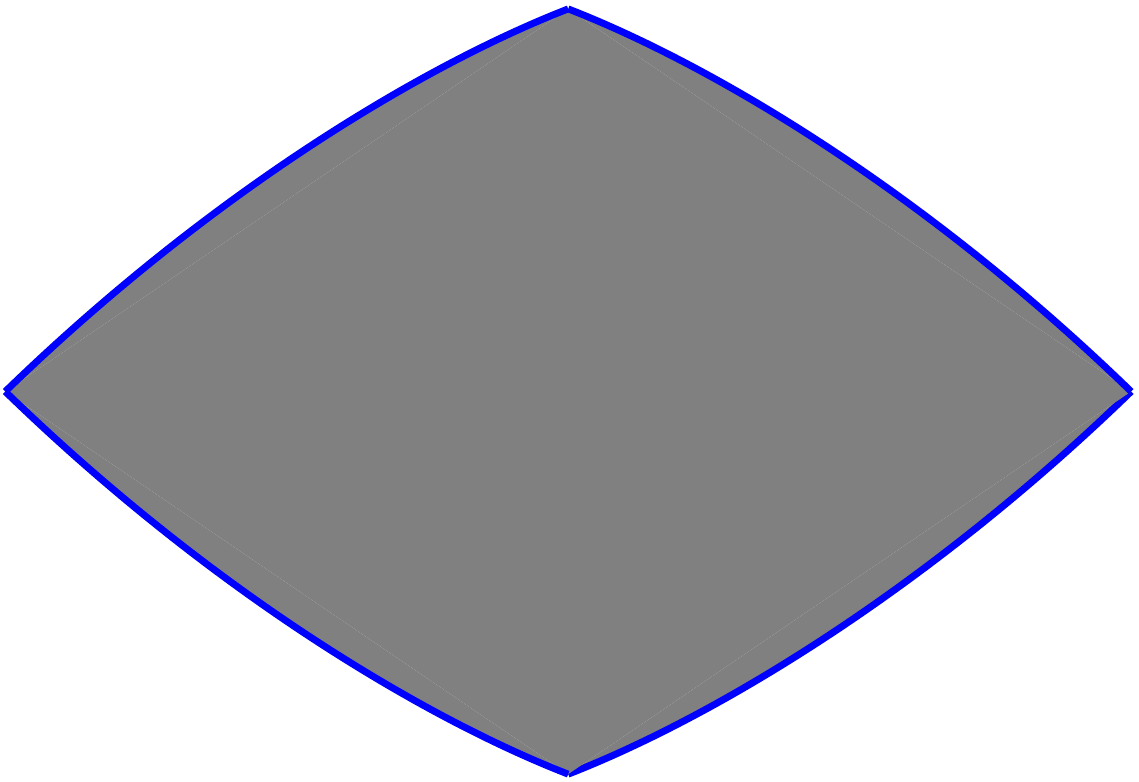}}
  \hspace{.05\linewidth}
   \subfigure[]{\includegraphics[width=.25\linewidth]{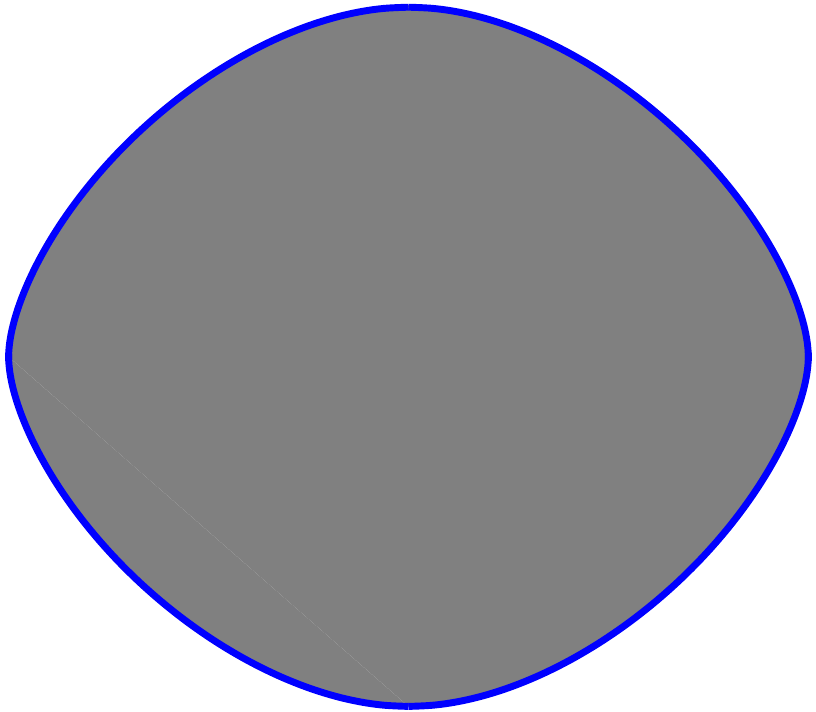}}
  \caption{
  Three (unscaled) equilibrium shapes for a \TBN drop. (a) Region I: $c=\frac{\sqrt{3}}{2}$, $\omega=\frac32$. (b) Region II: $c=\frac{\sqrt{3}}{2}$, $\omega=5$. (c) Region III: $c=\frac{\sqrt{3}}{2}$, $\omega=\frac47$.
  }
\label{fig:gallery}
\end{figure}

\begin{figure}[ht]
  \centering
  \subfigure[]{\includegraphics[width=.3\linewidth]{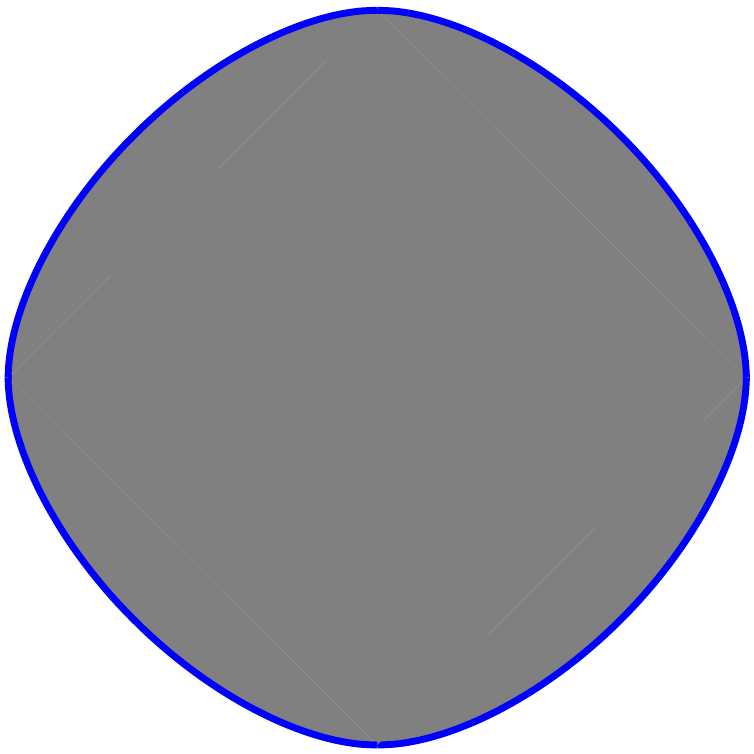}}
  \hspace{.05\linewidth}
  \subfigure[]{\includegraphics[width=.3\linewidth]{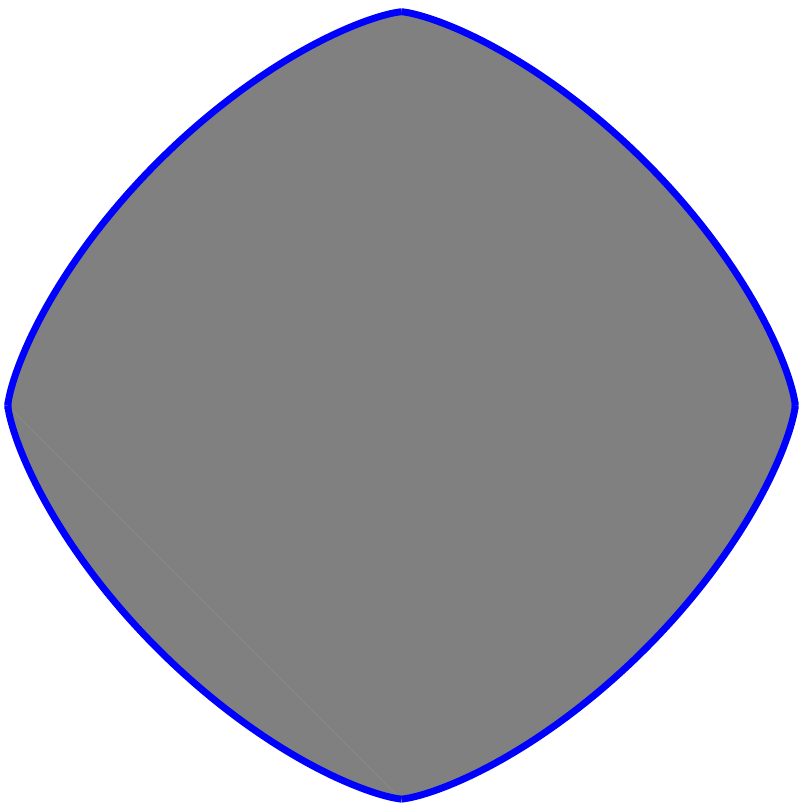}}
  \hspace{.05\linewidth}
  \subfigure[]{\includegraphics[width=.3\linewidth]{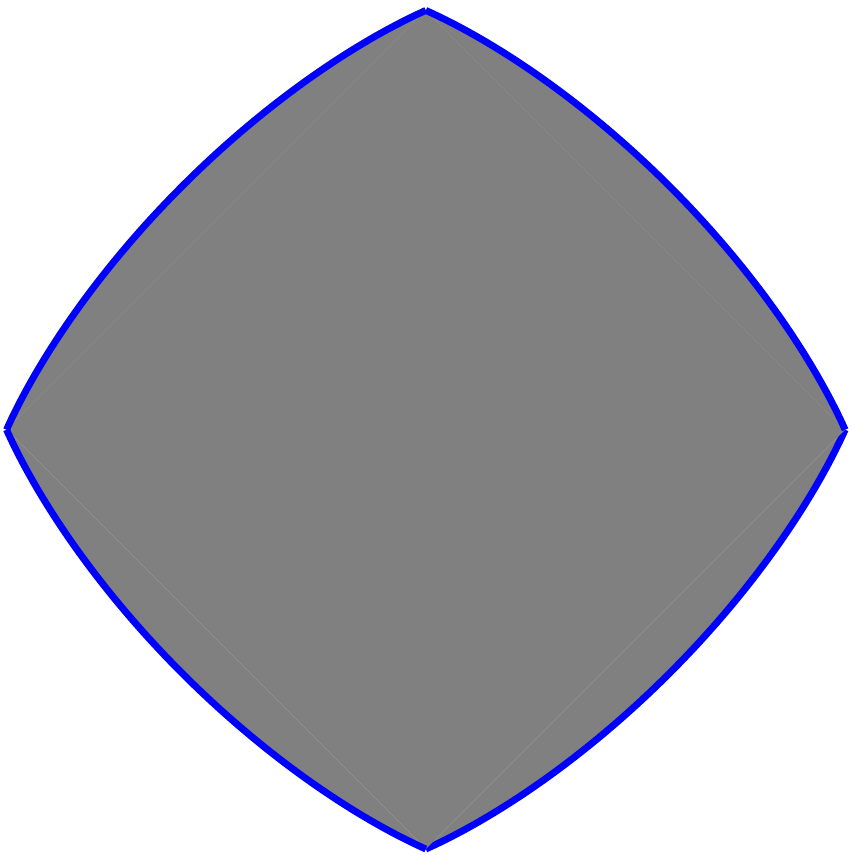}}
  \hspace{.05\linewidth}
  \subfigure[]{\includegraphics[width=.3\linewidth]{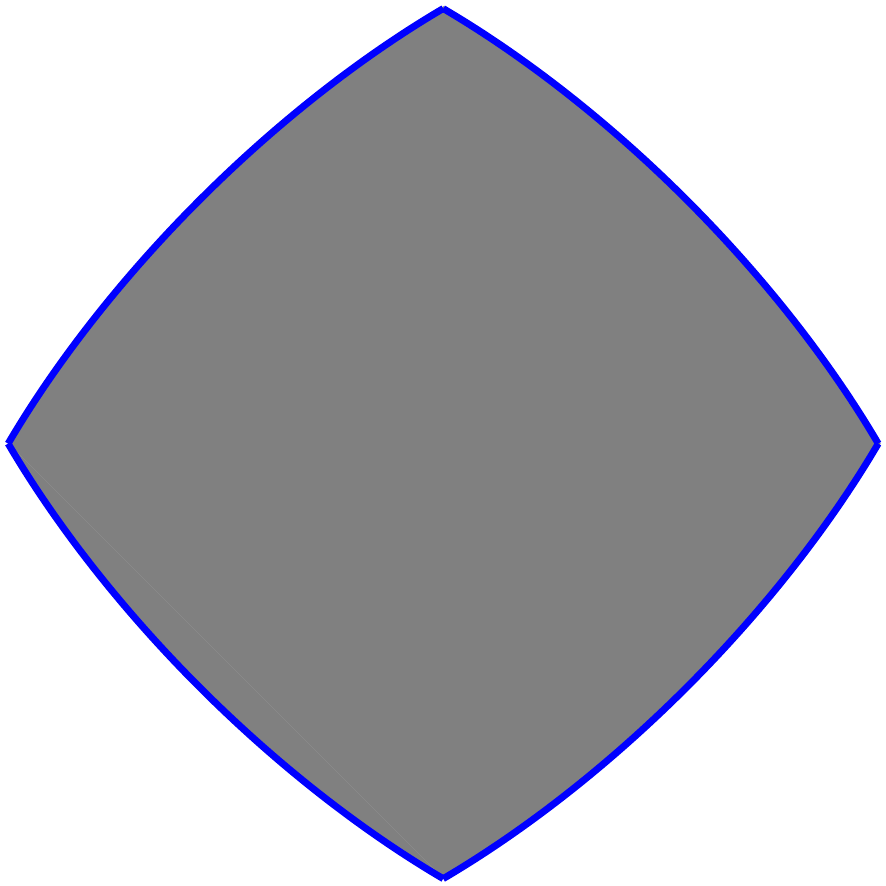}}
  \caption{
  Equilibrium shapes of a two-dimensional \TBN drop on the symmetry separatrix $c=\sepa$ of the parameter plane $(c,\omega)$. (a) $\omega=\frac47$; (b) $\omega=\frac87$; (c) $\omega=2$; (d) $\omega=3$. The shape (b) corresponds to the triple point $(\sepa,\frac87)$ of the phase diagram in Fig.~\ref{fig:phase_diagram}, where all regions I, I$'$, II, II$'$, III, and III$'$ meet.
  }
\label{fig:separatrix}
\end{figure}

We close this section by presenting in Fig.~\ref{fig:gallery} a gallery of equilibrium two-fold symmetric shapes for a \TBN drop in a uniformly aligned nematic phase.
They are to be scaled by an appropriate factor to enclose any desired area. All shapes shown were obtained via \eqref{eq:regular_equilibrium_arc_parametric_solution} for $\sepa\leqq c\leqq1$. They are all prolate along $\n$ as $\rho>1$; the mapping $c\mapsto\sqc$ rotates them by $\frac\pi2$ about the $z$ axis, exchanging the $x$ and $y$ axes and making them all oblate along $\n$. Figure~\ref{fig:separatrix} illustrates equilibrium shapes of the drop on the symmetry separatrix $c=\sepa$ of the parameter plane $(c,\omega)$. They also are all invariant under a $\frac\pi2$ rotation about the $z$ axis and have aspect ratio $\rho=1$.

\section{Conclusion}\label{sec:conclusion}
In this optical study on the morphological and interfacial
aspects of the \TBN phase in CB7CB, we have focussed on the
geometry of \TBN droplets in the environment of planarly aligned
host nematic monodomain sandwiched between two closely
spaced parallel plates. Well defined shapes are found in both
nonequilibrium and quasi-equilibrium situations. \TBN germs
nucleating immediately below their setting point grow
preferentially along the normal to the nematic director to result
in millimetric, near-elliptical droplets extending maximally
across $\n$. The droplet boundary is punctuated by cusplike
formations at the terminations of the principal axes. The
geometry here corresponds to a nonequilibrium situation in
which the anisotropy of thermal diffusivity seems more
dominant than the competing anisotropy of surface tension.

Equilibration of \TBN droplets is experimentally realized at a
temperature above the setting point, close to the melting
temperature. The drops then are tactoid-shaped, elongated
along the director of the surrounding nematic host. This
geometry is explained by invoking a mathematical model that
admits of an analytical solution, just as the classical Wulff
construction in 2D. The analytic jump conditions valid at an equilibrium corner played a major role in the model; they proved sufficient to draw a phase diagram for the drop, illuminating the qualitative features of all its possible equilibrium shapes (including the number of corners).

Combining experiment with theory and measuring both the aspect ratio and the inner corner angle of the observed cusped drops, we have extracted both parameters employed by the model, one of which is the ideal cone angle $\ca$ characteristic of the nanoscopic \TBN ground state. We estimate $\ca$ to be approximately $31.5\degree$, in good agreement with other structural measurements.\cite{meyer:temperature}

Other features were also revealed by our study, for which we cannot yet offer an equally satisfactory theoretical explanation. In particular, the micrometric quadrupole drop of dimensions comparable to the sample thickness, though qualitatively accounted for in our
phase diagram, seems to be more a dynamic transient state than a true equilibrium configuration. Another intriguing observation concerns the classical focal conic defect structure characteristic of \TBN phase at sufficiently low temperatures; it is observed to unravel progressively as the temperature gets closer to the $\mathrm{N_{tb}}$-N transition point. The intermediate structure---which for lack of a better description, is described here as the \emph{twin-striped} state---is not yet resolved, but we believe that it must be central to an understanding of the still mysterious $\mathrm{N_{tb}}$-N phase transition.

\section*{Acknowledgements}
We are thankful to Prof. O. D. Lavrentovich for fruitful discussions with one of us (EGV) and to Prof. G. U. Kulkarni for the experimental facilities.  One of us (CVY) acknowledges the financial support received from SERB, DST, Govt. of India, under Project No. SR/S1/OC-04/2012.

\appendix

\section{Analytical Details}
To arrive at a tractable form of the equilibrium equation in \eqref{eq:equilbrium_regular_arcs} for the regular arcs of $\curve$, we found it convenient to describe a quarter of a doubly symmetric curve $\curve$ as in Fig.~\ref{fig:quarter_drop}a in the form of a Cartesian graph $y=y(x)$. With $\curve$ thus reparameterized, the functional $F$ in \eqref{eq:F_functional} subject to the constraint of the area enclosed by $\curve$ can equivalently be rewritten as
\begin{equation}\label{eq:F_star_functional}
\begin{split}
&F^\ast[y]:=\\
&\int_0^a\left\{\left[1+\frac12\omega \left(\frac{1}{1+y'^2}-c^2\right)^2\right]\sqrt{1+y'^2}+\lambda y\right\}dx,
\end{split}
\end{equation}
which has also absorbed the area functional $A$ and the Lagrange multiplier $\lambda$ associated with it. In \eqref{eq:F_star_functional}, a prime now denotes differentiation with respect to $x$ and the function $y(x)$ is subject to
\begin{equation}\label{eq:y(a)=0}
y(a)=0,
\end{equation}
where $a>0$ is to be determined.

Since above we have already obtained the geometric conditions valid at the equilibrium corners of $\curve$, here we are only interested in finding the equilibrium arcs of $\curve$ in their Cartesian parametrization. The Euler-Lagrange equation associated with $F^\ast$ is easily obtained and integrated once, leading to
\begin{equation}\label{eq:F_star_E_L_equation}
\Phi(y';c,\omega)=\lambda x+b,
\end{equation}
where
\begin{equation}\label{eq:Phi_definition}
\begin{split}
\Phi(u;c,\omega):=\frac12\frac{1}{\sqrt{(1+u^2)^5}}\left[2-3\omega+2\omega c^2+\omega c^4\right.\\
\left. +2(2+\omega c^2+\omega c^4)u^2 +(2+\omega c^4)u^4 \right]u
\end{split}
\end{equation}
and $b$ is an arbitrary integration constant. Altering both $x$ and $y$ by the same factor, say $\mu$, so as to produce a homothetic dilation (or contraction) of $\curve$ does not alter the left side of \eqref{eq:F_star_E_L_equation}. Consequently, $\lambda$ must be changed into $\lambda/\mu$ for $\curve$ to remain an equilibrium curve. This shows that $\lambda$ can be determined (and the area constraint can be satisfied) by simply rescaling any solution $y=y(\xi)$ of \eqref{eq:F_star_E_L_equation} with $\xi:=\lambda x+b$. Similarly, since the differential equation \eqref{eq:F_star_E_L_equation} does contain $y$ explicitly, the constraint \eqref{eq:y(a)=0} can be satisfied by translating in space a solution.

\begin{figure}[ht]
  \centering
  \includegraphics[width=.9\linewidth]{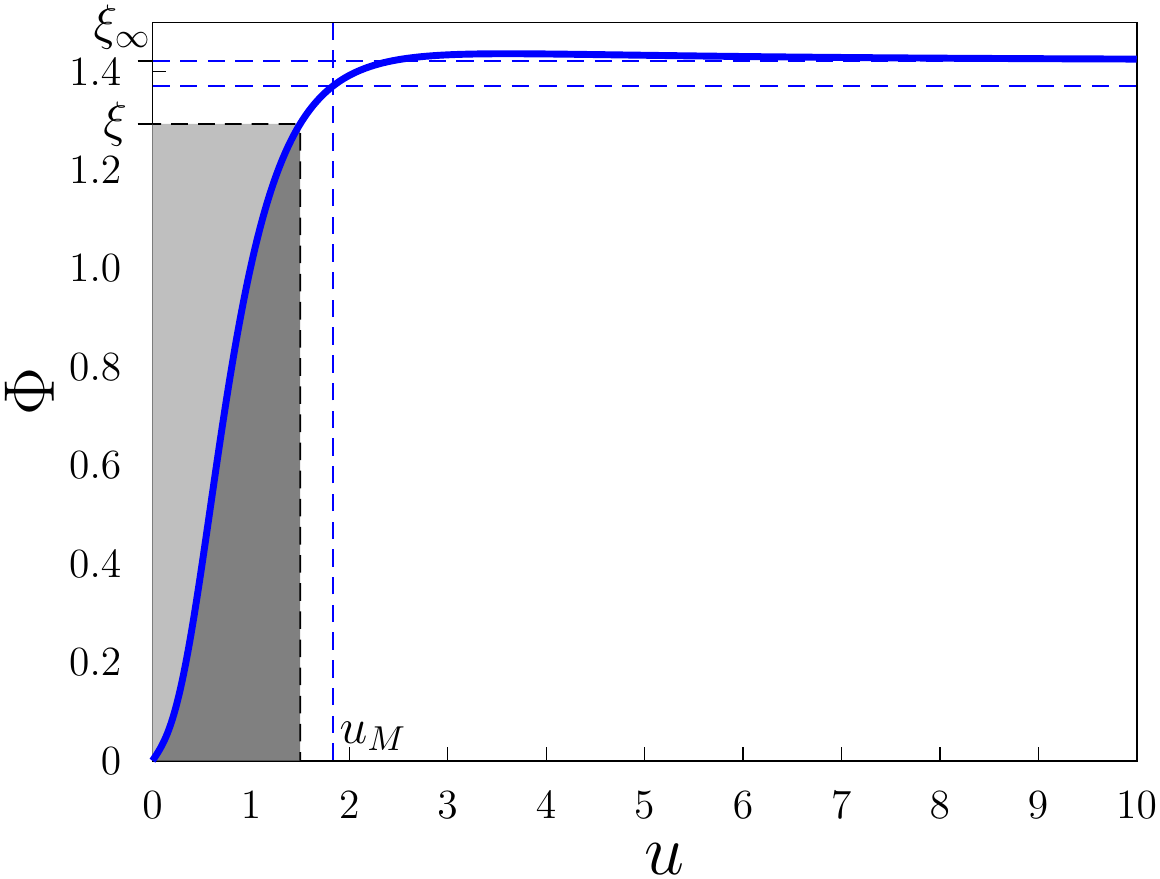}
  \caption{
  The plot of the function $\Phi$ against $u$ for $(c,\omega)$ in region I of the phase diagram in Fig.~\ref{fig:phase_diagram}. The asymptote $\xii$ is delivered by \eqref{eq:xi_infinity}; the single value $u_M$ of $y'$ where a regular arc of $\curve$ may meet a corner on the $x$ axis is identified by \eqref{eq:u_M}. This specific plot, while typical of the whole region I, was obtained for $c=\frac{\sqrt{3}}{2}$ and $\omega=\frac32$.}
\label{fig:Phi_One_bis}
\end{figure}

Figure~\ref{fig:Phi_One_bis} illustrates a graphical argument to integrate \eqref{eq:F_star_E_L_equation}.
For an arbitrary $\xi$, the light gray area represents the integral of $y'$ in $\xi$; such an area can be obtained by subtracting the dark gray area which represents the integral of $\Phi$ in $u$ from the whole area $u\Phi(u)$ of the rectangle delimited by the coordinate lines through $(u,\xi)$. Thus, to within additive constants to be chosen so as to adjust the solution to the geometric constraint, a regular arc of $\curve$ can be represented in the parametric form
\begin{subequations}\label{eq:regular_equilibrium_arc_parametric_solution}
\begin{align}
y &=Y(u;c,\omega):=u\Phi(u;c,\omega)-\Psi(u;c,\omega),\label{eq:y_parametric}\\
\xi &=\Phi(u;c,\omega),\label{eq:xi_parametric}
\end{align}
\end{subequations}
where $\Psi$ is the primitive of $\Phi$ in $u$. It is easily seen that for a regular arc with $x>0$ and $y<0$ as shown in Fig.~\ref{fig:quarter_drop}, $Y$ can be given the following explicit representation
\begin{equation}\label{eq:Y_representation}
\begin{split}
Y(u;c,\omega)=-\frac12\frac{1}{\sqrt{(1+u^2)^5}}[2+\omega-2\omega c^2+\omega c^4\\ +2(2+2\omega-3\omega c^2+\omega c^4)u^2 +(2-4\omega c^2+\omega c^4)u^4].
\end{split}
\end{equation}

With the aid of \eqref{eq:Phi_definition} and \eqref{eq:Y_representation}, by direct inspection one easily sees that the functions $\Phi$ and $Y$ enjoy the following property,
\begin{equation}\label{eq:shape_symmetry}
\Phi\left(\frac1u;\sqc,\omega\right)=Y(u;c,\omega),
\end{equation}
which combined with \eqref{eq:regular_equilibrium_arc_parametric_solution} mean that changing $c$ into $\sqc$ exchanges $\xi$ with $y$.

As shown in Fig.~\ref{fig:Phi_One_bis}, in region I a regular arc of $\curve$  can extend from $\xi=0$, where $y'=0$, to $\xi=\Phi(u_M)$, where $y'=u_M$, with $u_M$ related to $\chi_2$ in \eqref{eq:chi_2} through
\begin{equation}\label{eq:u_M}
u_M:=\sqrt{\frac{1}{\chi_2}-1}.
\end{equation}
At $\xi=\Phi(u_M)$, it meets an equilibrium corner on the $x$ axis, whereas it meets none on the $y$ axis. In principle, nothing would prevent one from further extending a regular arc for $\xi>\Phi(u_M)$, but as soon as $\xi$ crosses the asymptote of $\Phi$ at
\begin{equation}\label{eq:xi_infinity}
\xii:=1+\frac12\omega c^4,
\end{equation}
two distinct values of $y'$ can be associated with one and the same $\xi$ in equilibrium, a multiplicity compatible only with a corner of $\curve$ away from the symmetry axes, a case that here we have excluded from our consideration as we reckon it likely that these shapes  would be metastable. Thus, in region I all equilibrium shapes of $\curve$ considered here have two symmetric corners on the $x$ axis [see Fig.~\ref{fig:gallery}a above]; they are tactoids with axis along the direction of nematic alignment outside the drop.

\begin{figure}[ht]
  \centering
  \includegraphics[width=.9\linewidth]{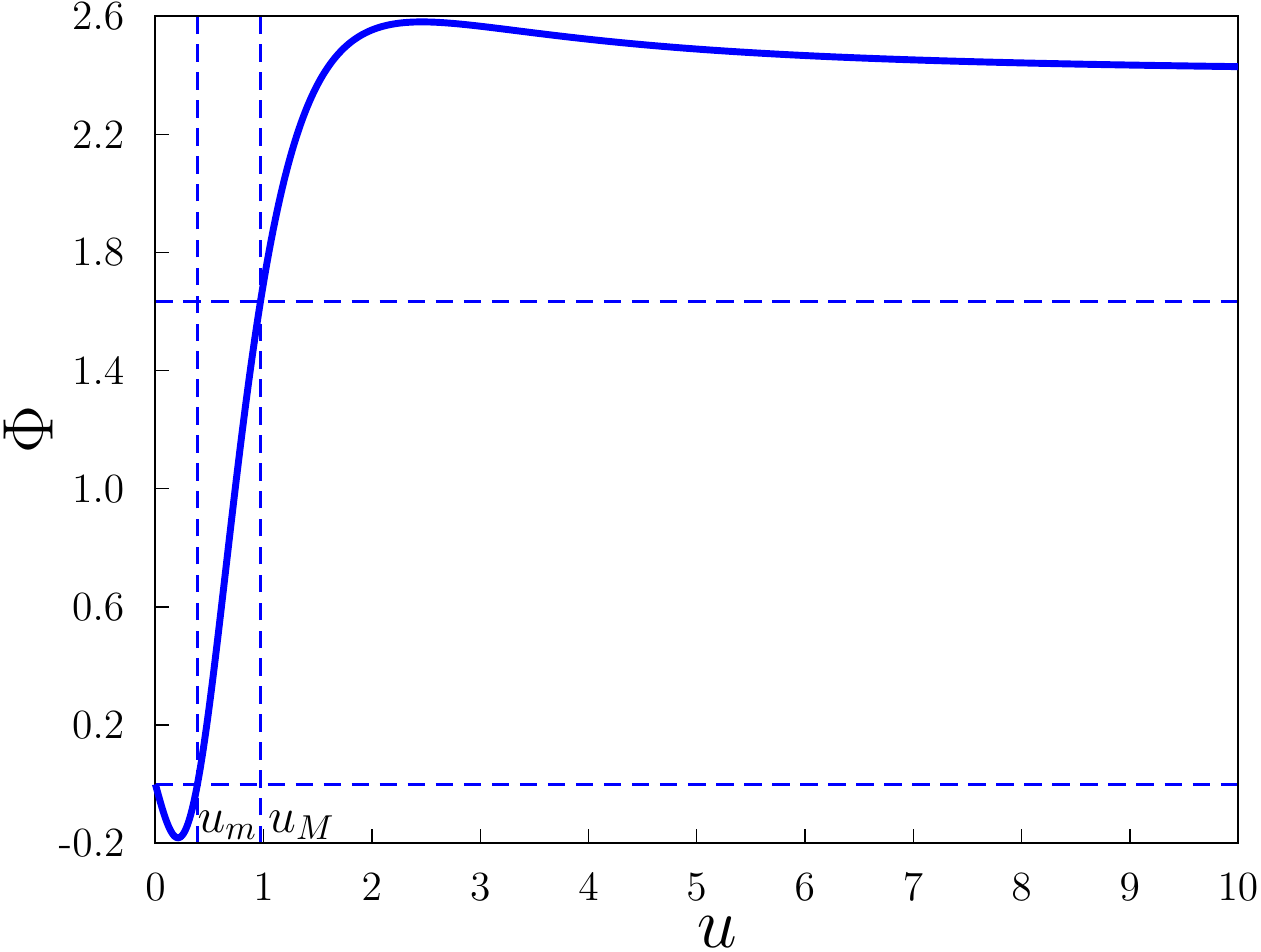}
  \caption{
  The plot of the function $\Phi$ against $u$ for $(c,\omega)$ in region II of the phase diagram in Fig.~\ref{fig:phase_diagram}. There are two values of $y'$, namely, $u_m$ and $u_M$ identified by \eqref{eq:u_m} and \eqref{eq:u_M}, where an equilibrium regular arc of $\curve$ can meet with corners on the two symmetry axes. Though characteristic of the whole region II, this specific graph was drawn for $c=\frac{\sqrt{3}}{2}$ and $\omega=5$.}
\label{fig:Phi_Four}
\end{figure}

In completely the same fashion, we analyze the equilibrium shape of $\curve$ in region II. Figure~\ref{fig:Phi_Four} illustrates the typical appearance of the graph of $\Phi$ in such a region.
Here $u_m$, which is related to $\chi_1$ in \eqref{eq:chi_1} through
\begin{equation}\label{eq:u_m}
u_m:=\sqrt{\frac{1}{\chi_1}-1},
\end{equation}
identifies the value of $y'$ where an equilibrium regular arc can meet a corner of the $y$ axis. These corners together with those met on the $x$ axis, where $y'=u_M$, make the equilibrium shape of the drop resemble a diamond [see Fig.~\ref{fig:gallery}b above]. Precisely, as above, one could try and extend an equilibrium regular arc of $\curve$ also for $y'<u_m$, but again the lack of monotonicity of $\Phi$ for $0<u<u_m$ is likely to bring $\curve$ in the realm of metastability.

\begin{figure}[ht]
  \centering
  \includegraphics[width=.9\linewidth]{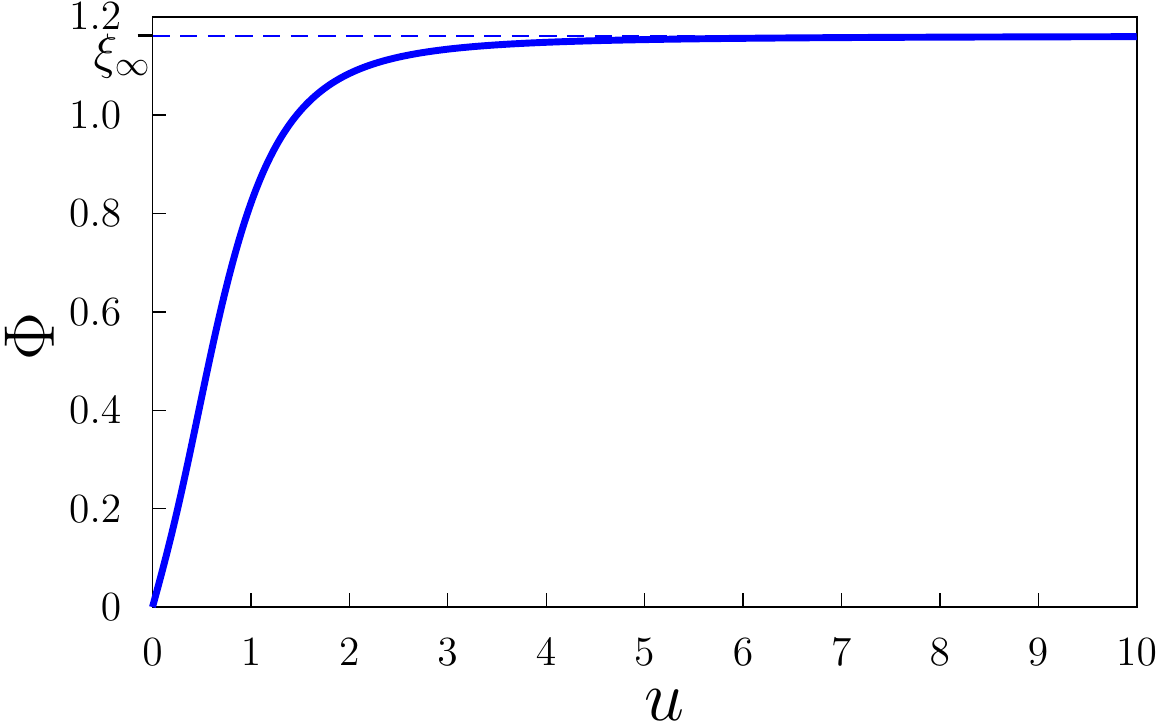}
  \caption{
  The plot of the function $\Phi$ against $u$ for $(c,\omega)$ in region III of the phase diagram in Fig.~\ref{fig:phase_diagram}. $\Phi$ is monotonically increasing in the whole range $u\geqq0$, saturating at $\xii$, still given by \eqref{eq:xi_infinity}.}
\label{fig:Phi_Three}
\end{figure}

Figure~\ref{fig:Phi_Three} illustrates the typical graph of $\Phi$ against $u$ in region III of the phase diagram in Fig.~\ref{fig:phase_diagram}.
Here $\Phi$ is monotonic and so $y'$ grows steadily from nought to infinity as $\xi$ traverses the interval $[0,\xii]$, with $\xii$ still given by \eqref{eq:xi_infinity}. The whole curve $\curve$ is smooth at equilibrium, as shown for example by Fig.~\ref{fig:gallery}c.

By \eqref{eq:regular_equilibrium_arc_parametric_solution}, for $\sepa\leqq c\leqq1$, the aspect ratio  $\rho$ of the extensions of the drop along $\n$ and orthogonally to $\n$ can be expressed as
\begin{equation}\label{eq:rho_formula}
\rho(c,\omega):=\left|\frac{\Phi(u_\mathrm{max};c,\omega)}{Y(u_\mathrm{min};c,\omega)}\right|,
\end{equation}
where
\begin{equation}\label{eq:u_min_u_max}
u_\mathrm{max}:=
\begin{cases}
u_M & \text{in}\ \mathrm{I}\cup\mathrm{II},\\
\infty & \text{in}\ \mathrm{III},
\end{cases}
\quad
u_\mathrm{min}:=
\begin{cases}
0 & \text{in}\ \mathrm{I}\cup\mathrm{III},\\
u_m & \text{in}\ \mathrm{II}.
\end{cases}
\end{equation}
As a consequence of \eqref{eq:shape_symmetry}, $\rho$ is easily seen to satisfy equation \eqref{eq:rho_symmetry} in the main text.





\balance
\providecommand*{\mcitethebibliography}{\thebibliography}
\csname @ifundefined\endcsname{endmcitethebibliography}
{\let\endmcitethebibliography\endthebibliography}{}

\end{document}